\documentclass[aps,prd,superscriptaddress,nofootinbib,11pt]{revtex4}
\usepackage[english]{babel}
\usepackage[utf8]{inputenc}
\usepackage{graphicx}   
\usepackage{slashed}
\usepackage{epstopdf}
\usepackage{verbatim}   
\usepackage{color}      
\usepackage{subfigure}  
\usepackage{multirow}
\usepackage{hyperref}   
\usepackage{float}
\usepackage{epsfig,rotating}
\usepackage{amsmath,amssymb}
\usepackage{dsfont}
\usepackage{slashed}
\restylefloat{table}
\numberwithin{equation}{section}

\newcommand{\vx}{\vec{x}}

\newcommand{\vk}{\vec{k}}

\newcommand{\be}{\begin{equation}}
\newcommand{\ee}{\end{equation}}
\newcommand{\bea}{\begin{eqnarray}}
\newcommand{\eea}{\end{eqnarray}}

\newcommand{\M}{\mathcal M}

\newcommand{\R}{\mathcal{R}}
\newcommand{\J}{\mathcal{J}}
\newcommand{\D}{\mathcal{D}}
\begin{document}
\title{Dynamics of axion-neutral pseudoscalar mixing.}

\author{Shuyang Cao}
\email{shuyang.cao@pitt.edu} \affiliation{Department of Physics, University of Pittsburgh, Pittsburgh, PA 15260, USA.}
\author{Wenjie Huang}
\email{weh68@pitt.edu} \affiliation{Department of Physics, University of Pittsburgh, Pittsburgh, PA 15260, USA.}
\author{Daniel Boyanovsky}
\email{boyan@pitt.edu} \affiliation{Department of Physics, University of Pittsburgh, Pittsburgh, PA 15260, USA.}

 \date{\today}

\begin{abstract}

Axions mix with neutral pions after the QCD phase transition through their common coupling to the radiation bath via a Chern-Simons term,  as  a consequence of the $U(1)$ anomaly.    The non-equilibrium effective action that describes this mixing phenomenon  is obtained to second order in the coupling of neutral pions and axions to photons. We show that a misaligned axion condensate induces a neutral pion condensate after the QCD phase transition. We argue that even a very small axion-photon coupling will ``seed'' the chiral phase transition. The dynamics of the pion condensate displays   long and short time scales and decays on the longer time scale exhibiting a phenomenon akin to the ``purification'' in a Kaon beam. On the intermediate time scales the macroscopic pion condensate is proportional to a condensate of the abelian Chern-Simons term induced by the axion. We  argue that the coupling to the common bath also induces kinetic mixing.  We obtain the axion and pion populations, and these exhibit thermalization with the bath. The mutual coupling to the bath induces long-lived axion- neutral pion coherence independent of initial conditions.   The framework of the effective action and many of the consequences are more broadly general and applicable to   scalar or pseudoscalar particles mixing in a medium.

\end{abstract}

\keywords{}

\maketitle

\section{Introduction}

The axion is a CP-odd (pseudo) scalar particle which was proposed as an extension beyond the standard model as a possible solution    of the strong CP problem in Quantum Chromodynamics (QCD)\cite{PQ,weinaxion,wil}. Such particle  may be produced non-thermally in the Early Universe,   and  has been  recognized as a compelling  cold dark matter candidate\cite{pres,abbott,dine}. Extensions beyond the standard model can accommodate pseudoscalar particles with properties similar to the QCD axion, namely axion-like-particles (ALP) which can also be suitable dark matter candidates\cite{banks,ringwald,marsh,sikivie1,sikivie2}, in particular as candidates for ultra light dark matter\cite{fuzzy,uldm}. Constraints on the mass and couplings of ultra light (ALP)\cite{marsh,sikivie1,sikivie2,banik} are being established  by various experiments\cite{cast,admx,graham}.  There are various possible production mechanisms for axion production available in the early Universe\cite{marsh,sikivie1,sikivie2}, among them a non-thermal  misalignment mechanism which results in an initial non-vanishing expectation value of the (ALP) field, which consequently evolves with coherent oscillations yielding a contribution to the energy density as a cold dark matter component\cite{pres,abbott,dine,marsh,sikivie1,sikivie2,turner}. Such misalignment mechanism may be a consequence of non-perturbative effects during the QCD phase transition.  Through the Peccei-Quinn\cite{PQ} U(1) anomaly the  pseudoscalar nature leads to an interaction between the (ALP) and photons  or  gluons via pseudoscalar composite operators of gauge fields, namely Chern-Simons terms  such as $\vec{E}\cdot\vec{B}$ in the case of the (ALP)-photon interaction and $G^{\mu\nu;b}\widetilde{G}_{\mu\nu;b}$ in the case of gluons, which allows an (ALP) to decay into two photons or gluons,  effects of  such decay process on the (ALP)  evolution have been studied in refs.\cite{sigl,arza,dashin}. More recently the non-equilibrium effective action of (ALP) interacting with photons in a radiation bath has been derived\cite{shuyang}, and from which the effective equations of motion for the (ALP) condensate were obtained. These are stochastic equations of the Langevin type, with friction and noise kernels that fulfill a generalized fluctuation dissipation relation\cite{boyqbm}. The solutions of this Langevin equation  allowed us to obtain the energy density which revealed the approach to thermalization with the cosmic microwave background, implying in turn, a mixed cold and hot dark matter scenario. These results were confirmed via an alternative method based on the quantum master equation in ref.\cite{shuyangqme}.
 Furthermore, it was shown in ref.\cite{shuyangcs} that an expectation value of the axion field, namely a ``misaligned'' axion, can induce an expectation value of the Chern-Simons density. A corollary of this result is that neutral $\pi^0$ mesons, which also couple to photons via the U(1) anomaly,   mix with axions via this common two photon channel. As a result of this indirect mixing,  misaligned axions   induce an expectation value of the neutral meson field.

Axions may also play a role in condensed matter physics, possibly  as emergent quasiparticles in  topological insulators where magnetic fluctuations couple to electromagnetism just like axions\cite{wilczekaxion,wang,narang,nomura}, as axionic charge density waves in Weyl semimetals\cite{gooth,yu},   as an emergent axion response in multilayered metamaterials with tunable couplings\cite{wilczek} or in multiferroics\cite{baly}. The measurement of an emergent dynamic axion field in chromia has been reported in ref.\cite{binek},  therefore, condensed matter systems may very well provide an experimental platform to test the main aspects of axion electrodynamics which may complement and bolster the case for axions in cosmology. Hence,  the study of axion (electro) dynamics is of timely interdisciplinary relevance bridging cosmology and condensed matter with the tantalizing possibility of probing the cosmological axion with novel   condensed matter systems.

\textbf{Motivation and objectives:}
The Peccei-Quinn  axion has been  motivated as a solution of the strong CP problem in QCD, and the misalignment mechanism yielding the axion condensate making it a compelling cold dark matter candidate may be a consequence of the QCD phase transition. Such phase transition has another important consequence: the binding of the lightest $u,d$ quarks   into the lightest pseudoscalar isotriplet of $\pi$ mesons, namely    pions, which are copiously produced at the QCD phase transition at $T_{QCD} \simeq 150-170\,\mathrm{MeV} \geq m_{\pi} \simeq 140\,\mathrm{MeV}$. Among the three pion states $\pi^0,\pi^\pm$, the neutral pion $\pi^0$ couples to the electromagnetic field via the U(1) triangle anomaly, resulting in that it decays into two photons with a $\simeq 99\%$ branching ratio. The neutral pion has the same quantum numbers as the axion: it is a neutral pseudoscalar, and just as the axion its coupling to electromagnetism is a consequence of the $U(1)$ triangle anomaly.

Several extensions beyond the Standard Model can accommodate axion-like particles, however, a common thread among these hypothetical particles is that they couple to photons and gluons via a Chern-simons term, thereby allowing a two-photon decay channel, thus generalizing the original proposal for the QCD axion.

The main premise of this study is that the neutral pion and the axion \emph{mix} via a common set of two-photons intermediate states: consider the processes $\pi^0 \leftrightarrow 2\gamma\leftrightarrow a$ with $a$ the axion, such process induces a mixing between the two fields in the form of an off-diagonal self-energy. Such mixing is enhanced in the thermal medium at the temperature of the QCD phase transition since photons in the bath feature thermal energies sufficient to produce $\pi^0$ in detailed balance with the annihilation (decay) process. This mixing mechanism via a common decay channel for neutral pions and axions was anticipated in  ref.\cite{shuyangcs} where it was suggested that   as a consequence of such mixing, a misaligned axion condensate will induce a neutral pion condensate.

Our motivation in this study stems from this observation, namely  that the axion will mix with the neutral pion after the QCD phase transition through a common intermediate state of two photons in the medium. In fact, this observation leads to a more overarching possibility: if there are other axion-like particles emerging from extensions beyond the standard model that, just like the QCD axion, couple to photons or gluons as a consequence of a gauge anomaly, these alternative  (ALP)'s will also couple to the QCD axion via a common intermediate state of   photons or gluons.

Therefore, beyond the QCD axion, other possible (ALP), will mix among themselves as well as with the neutral pion after the QCD phase transition via their mutual coupling to photons and or gluons through the Chern-Simons density.

However, studying the dynamics of field mixing in a medium is of much broader  fundamental and intrinsic interest  within the context of CP violation and/or baryogenesis. A paradigmatic example in \emph{vacuum} is the mixing of $K^0-\overline{K}^0$ or flavored meson-antimesons as a consequence of a common intermediate state, which provide dynamical observational signatures of CP violation\cite{sanda,nierste,pila}. Field mixing may also be a consequence of  ``portals'', connecting standard model degrees of freedom to hypothetical ones beyond the standard model,  such ``portals'' may lead to mixing between fields on different sides of the ``portal''. Hence the  study of the dynamics of mixing in a medium may open hitherto unexplored new windows into  extensions beyond the standard model.

Furthermore, axion electrodynamics in topological materials and/or Weyl semimetals also features a  coupling of these emergent collective excitations to electromagnetism via processes akin to the U(1) anomaly\cite{wilczekaxion,wang,narang,nomura,gooth,yu,wilczek,binek}. Therefore, these ``synthetic'' axions may mix with the cosmological axion in the same manner as pions or generic (ALP)'s do in the early Universe, hence  motivating the study of the possibility of mixing between the cosmological and the ``synthetic'' axions. Such mixing may yield   alternative insights into probing cosmological axions with condensed matter experiments.

Our objectives in this study are two-fold: \textbf{I:)} to develop the framework to study the  \emph{dynamics} of scalar or pseudoscalar field mixing in a medium via a common intermediate channel corresponding to degrees of freedom in thermal equilibrium. This indirect mixing is a consequence of off diagonal contributions to the self-energy where the intermediate states (in the loops) are those corresponding to the bath degrees of freedom.  This is achieved by obtaining the Schwinger-Keldysh (in-in)
non-equilibrium effective action\cite{schwinger,keldysh,maha,feyver,beilok} of the mixing degrees of freedom after tracing over the bath degrees of freedom,  by extending and generalizing the formulation of the non-equilibrium effective action\cite{boyqbm}, and applied to   axion dynamics   in ref.\cite{shuyang}. The framework developed within this study will find a natural arena in cosmology.

\textbf{II:)} To implement this general framework in the specific example of axion-neutral pion mixing after the QCD phase transition. There are several reasons for studying this particular case: \textbf{a:)} it provides an explicit realization of the main concepts behind indirect mixing via a common set of intermediate states in terms of off-diagonal self-energies, \textbf{b:)} (ALP) mixing, and or (ALP)-neutral pion mixing are of cosmological interest, \textbf{c:)} the operator that describes the common set of intermediate states is clearly identified and amenable of a spectral analysis,  \textbf{d:)} it allows us to scrutinize one of the conjectures of ref.\cite{shuyangcs}, namely  that a misaligned axion condensate will induce a neutral pion condensate, which in turn is related to a Chern-Simons condensate\cite{shuyangcs}. As it will become clear, this example provides the setting for a wide separation of time scales, leading to decaying eigenstates with short and long lifetimes, akin to (albeit  with important differences with) the kaon system,  and  illuminating fundamental aspects related to the dynamics of mixing, such as  the emergence of long-lived correlations and ``purification''. Furthermore, we show that a misaligned axion \emph{seeds} a pion condensate, and we argue that  even for a very small axion-photon coupling, the ``seed'' of pion condensate will be amplified by the inherent instabilities associated with the chiral phase transition.

\vspace{1mm}

\textbf{Brief summary of results:} \textbf{a:)} We obtain the non-equilibrium effective action for scalar or pseudoscalar mixing indirectly through their mutual coupling to  a common bath.  The equations of motion are stochastic with dissipative and noise kernels obeying generalized fluctuation dissipation. These are solved for the general case and applied to axion-neutral pion mixing after the QCD phase transition. \textbf{b:)} A misaligned axion condensate induces a macroscopic neutral pion condensate that ``seeds'' the chiral phase transition. The induced pion condensate   displays time evolution on a long and a short time scale, akin to the $K^0-\overline{K}^0$ in vacuum (albeit with important differences). The pion condensate survives on time scales much longer than the pion decay lifetime, a phenomenon reminiscent of the ``purification'' of a Kaon beam. \textbf{c:)} On time scales much longer than the pion lifetime, the neutral pion condensate is proportional  to the macroscopic U(1) Chern-Simons condensate. \textbf{d:)} Indirect mixing also implies the necessity for kinetic mixing. \textbf{e:)} The axion and pion populations exhibit thermalization with the radiation bath, while connected correlation functions exhibit long-lived off-diagonal coherence, a manifestation of ``bath-induced coherence''.

In section (\ref{sec:effact}), we introduce the generic model to be studied,   obtain the Schwinger-Keldysh (in-in) effective action, and derive  and solve the equations of motion of the mixing fields in the general case. In this section we discuss subtle renormalization aspects and the necessity for kinetic mixing in the effective action. Section (\ref{sec:piondynamics}) is devoted to studying the dynamics of axion-neutral pion mixing as a specific example, from which we draw more general lessons. In this section we confirm one of the main conjectures of ref.\cite{shuyangcs}, namely that a misaligned axion induces a neutral pion condensate via mixing, and study the time evolution of correlations, coherences and populations.
Section (\ref{sec:conclusions}) summarizes our results and conclusions. Several appendices provide technical details.

\section{The Non-Equilibrium Effective Action}\label{sec:effact}

To set the stage for the general formulation and to highlight the main concepts behind indirect mixing in a medium, let us consider the following Lagrangian density describing axions and neutral pions $a,\pi^0$ respectively coupled to a photon bath via a Chern-Simons term,  as an effective field theory description of the system to be studied,
\be \mathcal{L} = \frac{1}{2}\,\partial_{\mu}\pi^0\partial^\mu \pi^0-\frac{1}{2} m^2_\pi\,{\pi^0}^2-\frac{\alpha}{8\pi\,f_\pi}\,\pi^0\,F_{\mu\nu}\widetilde{F}^{\mu \nu}+ \frac{1}{2}\,\partial_{\mu}a\partial^\mu a-\frac{1}{2} m^2_a\,a^2-\frac{g_a}{4}\,a\,F_{\mu\nu}\widetilde{F}^{\mu \nu}+\mathcal{L}_\gamma \,, \label{axpilag}\ee    where $\widetilde{F}^{\mu \nu} = \frac{1}{2} \, \epsilon^{\mu\nu\rho\sigma}\,F_{\rho\sigma}$.    $\alpha$ is the fine structure constant and  $f_\pi$ the pion decay constant together yield  $\alpha/\pi f_\pi \simeq 0.025\,\mathrm{GeV}^{-1}$, for QCD with three quark colors, as determined by the neutral pion decay width into two photons. The axion coupling is $g_a = \mathcal{C}/f_a$ with $\mathcal{C}$ a dimensionless constant and $f_a$ the axion decay constant. In the effective Lagrangian (\ref{axpilag}), $\mathcal{L}_\gamma$ is the Lagrangian density for the electromagnetic fields, including coupling to charged degrees of freedom within (or beyond) the Standard Model. At finite temperature both the pion mass and decay constant $f_\pi$ are modified by   thermal corrections differing by about $15\%$ from the zero temperature values\cite{piontem}. However, in this study we will not be concerned with the actual values of the couplings and masses, focusing instead on the fundamental aspects of the dynamics of mixing and its   consequences. Both the axion and neutral pion couple to the radiation bath of photons in equilibrium via the $U(1)$   Chern-Simons density $-F_{\mu\nu}\widetilde{F}^{\mu \nu}/4 = \vec{E}\cdot\vec{B}$. Hence both can decay into two photons or absorb photons from the bath, thereby allowing photons to mediate axion-pion mixing via the processes $\pi^0 \leftrightarrow 2\gamma \leftrightarrow a$. These processes yield an off-diagonal contribution to the axion and pion self-energies (see fig. (\ref{fig:pionaxion}) below). This is the main concept underpinning ``indirect'' mixing\footnote{Direct mixing would correspond to an off-diagonal mass matrix for example.}.

Since we seek to understand the mixing of axions with pions, as well as with other possible (ALP) candidates,   we generalize the effective Lagrangian (\ref{axpilag}) to describe two different pseudoscalar fields $\phi_{1,2}$, of masses $m_{1,2}$ respectively coupled to a common bath of photons in thermal equilibrium via the Chern-Simons density.   The radiation bath is itself coupled to and in thermal equilibrium with other charged degrees of freedom,  denoting these bath degrees of freedom generically  as $\chi$. Hence, we propose  to study mixing dynamics described by the generalized Lagrangian density given by

\begin{equation}
    \mathcal{L}[\phi_1,\phi_2,\chi] = \mathcal{L}_\phi + \mathcal{L}_{\chi} + \mathcal{L}_I \,,\label{lag}
\end{equation}
where
\bea
    \mathcal{L}_\phi
    & = & \frac{1}{2}\,\sum_{a=1,2}     \,  \big(\partial \phi_a\big)^2 - \frac{1}{2} m^2_a\, \phi^2_a-\frac{1}{2}\sum_{a,b=1,2}\phi_a\,\Delta m^2_{ab}\phi_b      \nonumber    \\
    \mathcal{L}_I
    & = &    (g_1\,\phi_1 +g_2\, \phi_2) \,\mathcal{O}[\chi]\,,\label{lagra} ~~;~~ \mathcal{O}[\chi]\equiv \vec{E}\cdot\vec{B} \,.\label{genlag}
\eea
 Anticipating renormalization we introduced a  matrix of mass counterterms $\Delta m^2_{ab}$ required to cancel the ultraviolet divergences in the self-energies.   $\mathcal{L}_\chi$ is the Lagrangian of electromagnetic fields, including coupling to other charged degrees of freedom,  assumed  to describe the bath degrees of freedom in thermal equilibrium. The general form   (\ref{genlag}) allows us to treat the different cases within one framework, from which we can extract individual cases by specifying masses and couplings. In what follows we will always consider $\phi_2$ as the (QCD) axion field, with $\phi_1$ being either the neutral pion, or alternative (ALP) fields.

 We shall first obtain the non-equilibrium effective action for general masses and couplings by extending the methods introduced in refs.\cite{shuyang,boyqbm}, addressing  each specific case a posteriori. In $3+1$ space-time dimensions, the composite operator $\vec{E}\cdot\vec{B}$ features mass dimension four, therefore the couplings $g_{1,2}$ feature dimensions $(\mathrm{mass})^{-1}$, and the theory defined by the Lagrangian density (\ref{genlag}) is non-renormalizable. It must be understood as an effective field theory valid below a scale $\Lambda$, which for consistency will be taken to satisfy $\Lambda \gg m_{1,2},T$, where $T$ is the temperature of the bath degrees of freedom. The steps towards obtaining the non-equilibrium effective action are the following\cite{shuyang,boyqbm}: an initial density matrix describing the fields $\phi_{1,2}$ and the bath degrees of freedom $\chi$ is evolved in time with the unitary time evolution operator, a trace over the bath degrees of freedom $\chi$ yields a \emph{reduced} density matrix associated solely with the fields $\phi_{1,2}$, this reduced density matrix \emph{does not} evolve unitarily in time. The non-equilibrium effective action yields the   kernel that determines the time evolution of the reduced density matrix and includes the dissipative effects associated with tracing out the $\chi$ degrees of freedom in terms of self-energies. These steps are implemented in detail below.

\subsection{The non-equilibrium effective action:}\label{subsec:noneq}

  Let us consider the initial density matrix at a time
$t=0$ to be of the form
\begin{equation}
\hat{\rho}(0) = \hat{\rho}_{\phi}(0) \otimes
\hat{\rho}_{\chi}(0) \,.\label{inidensmtx}
\end{equation}

The initial density matrix $\hat{\rho}_\phi(0)$ is normalized so that  $\mathrm{Tr}_\phi \hat{\rho}_\phi(0) =1$ and that of the $\chi$ fields will be taken to
describe a statistical ensemble in thermal equilibrium at a temperature
$T=1/\beta$, namely

\begin{equation}\label{rhochi}
\hat{\rho}_{\chi}(0) = \frac{e^{-\beta\,H_{\chi}}}{\mathrm{Tr}_{\chi} e^{-\beta H_\chi}}\,,
\end{equation}

\noindent where $H_{\chi} $ is the total Hamiltonian for the $\chi$ degrees of freedom.

The factorization of the initial density matrix is an assumption often explicitly or implicitly made in the literature, it may be relaxed by including initial correlations between the   fields $\phi_{a}$ and the bath fields $\chi$ at the expense of daunting technical complications. We will not consider   this important case, relegating it to future study. In what follows we will refer to the set of fields $\phi_{1,2}$ collectively simply as $\phi$ to simplify notation.

  In the field basis the matrix elements of $\hat{\rho}_{\phi}(0)$ and $\hat{\rho}_{\chi}(0)$
are given by
\begin{equation}
\langle \phi |\hat{\rho}_{\phi}(0) | \phi'\rangle =
\rho_{\phi,0}(\phi ,\phi')~~;~~\langle \chi|\hat{\rho}_{\chi}(0) | \chi'\rangle =
\rho_{\chi,0}(\chi ;\chi')\,,
\end{equation} we emphasize that this is a \emph{functional} density matrix as the fields have spatial arguments.
The density matrix for the   fields $ \phi$   represents either  a pure state or more generally an initial
 statistical ensemble, whereas $\hat{\rho}_{\chi}(0)$ is given by eqn. (\ref{rhochi}).

  To obtain the effective action  out of equilibrium for the fields $\phi_{1,2}$ (describing the (ALP) and neutral pion fields in (\ref{axpilag})),   we   evolve the initial density matrix in time and trace over the ``bath'' degrees of freedom, leading to a reduced density matrix for the   fields $\phi_{1,2}$, from which
 we can compute its expectation values or correlation functions as a function of time.

The time evolution of the   density matrix in the Schroedinger picture is given by

\begin{equation}
\hat{\rho}(t)= U(t)\hat{\rho}(0)U^{-1}(t)\,, \label{rhooft}
\end{equation}
where
\begin{equation}
U(t) = e^{-iHt}\,. \label{unitimeop} \ee
The total Hamiltonian $H$ is given by
\begin{equation}
H=H_{0 \phi} + H_{\chi}+ \int d^3x \sum_{a=1,2} g_a\,\phi_{a}\, \mathcal{O} [\chi]  \,, \label{hami}
\end{equation}
and $H_{0 \phi},H_{\chi}$  are the   Hamiltonians for   the respective fields.

The \emph{reduced} density matrix for the  $\phi_{1,2}$ fields is obtained by tracing over the $\chi$ degrees of freedom as
\be \rho^{r}_{\phi}(t) = \mathrm{Tr}_{\chi} U(t) \hat{\rho}(0)\,U^{-1}(t)\,.  \label{rored}\ee To extract the non-equilibrium effective action   it is more convenient to obtain the density matrix elements in field space, namely

\be    \rho(\phi_f,\chi_f;\phi'_f,\chi'_f;t)   =        \langle \phi_f;\chi_f|U(t)\hat{\rho}(0)U^{-1}(t)|\phi'_f;\chi'_f\rangle \,,\label{evolrho} \ee from which the reduced density matrix elements are
\be \rho^r(\phi_f ;\phi'_f,;t) = \int D\chi_f \,\langle \phi_f;\chi_f|U(t)\hat{\rho}(0)U^{-1}(t)|\phi'_f;\chi_f\rangle \,. \label{redmtxel}\ee

With the functional integral representation
\bea \langle \phi_f;\chi_f|U(t)\hat{\rho}(0)U^{-1}(t)|\phi'_f;\chi'_f\rangle  & = &  \int D\phi_i D\chi_i D\phi'_i D\chi'_i ~ \langle \phi_f;\chi_f|U(t)|\phi_i;\chi_i\rangle\,\rho_{\phi,0}(\phi_i;\phi'_i)\, \otimes   \nonumber \\ & & \rho_{\chi,0}(\chi_i;\chi'_i) \,
 \langle \phi'_i;\chi'_i|U^{-1}(t)|\phi'_f;\chi'_f\rangle \,,\label{evolrhot}\eea from which it follows that the reduced
 density matrix elements are
\bea \rho^r(\phi_f ;\phi'_f,;t) & = & \int D\chi_f   \int D\phi_i D\chi_i D\phi'_i D\chi'_i ~ \langle \phi_f;\chi_f|U(t)|\phi_i;\chi_i\rangle\,\rho_{\phi,0}(\phi_i;\phi'_i)\, \otimes   \nonumber \\ & & \rho_{\chi,0}(\chi_i;\chi'_i) \,
 \langle \phi'_i;\chi'_i|U^{-1}(t)|\phi'_f;\chi_f\rangle \,.\label{redfun} \eea

  The $\int D\phi$ etc, are functional integrals where the spatial argument has been suppressed. The matrix elements of the time evolution forward and backward can be written as path integrals, namely
 \bea   \langle \phi_f;\chi_f|U(t)|\phi_i;\chi_i\rangle  & = &    \int \mathcal{D}\phi^+ \mathcal{D}\chi^+\, e^{i \int d^4 x \mathcal{L}[\phi^+,\chi^+]}\,,\label{piforward}\\
 \langle \phi'_i;\chi'_i|U^{-1}(t)|\phi'_f;\chi'_f\rangle &  =  &   \int \mathcal{D}\phi^- \mathcal{D}\chi^-\, e^{-i \int d^4 x \mathcal{L}[\phi^-,\chi^-]}\,,\label{piback}
 \eea where we use the shorthand notation
 \be \int d^4 x \equiv \int_0^t dt' \int d^3 x \,.\label{d4xdef}\ee
 $ \mathcal{L}[\phi,\chi] $ is given by (\ref{lag},\ref{lagra})   and
 the boundary conditions on the path integrals are
  \bea     \phi^+(\vec{x},t=0)=\phi_i(\vec{x})~;~
 \phi^+(\vec{x},t)  &  =  &   \phi_f(\vec{x})\,,\label{piforwardbc}\\   \chi^+(\vec{x},t=0)=\chi_i(\vec{x})~;~
 \chi^+(\vec{x},t) & = & \chi_f(\vec{x}) \,,\label{chipfbc}  \\
     \phi^-(\vec{x},t=0)=\phi'_i(\vec{x})~;~
 \phi^-(\vec{x},t) &  = &    \phi'_f(\vec{x})\,,\label{aminbc} \\   \chi^-(\vec{x},t=0)=\chi'_i(\vec{x})~;~
 \chi^-(\vec{x},t) & = & \chi'_f(\vec{x}) \,.\label{pibackbc}
 \eea

The field variables $\phi_a^\pm, \chi^\pm$ along the forward ($+$) and backward ($-$) evolution branches are recognized as those necessary for the in-in or  Schwinger-Keldysh\cite{schwinger,keldysh,maha,beilok} closed time path approach to the time evolution of a density matrix.

 The reduced density matrix for the  fields $\phi_a$ (\ref{redfun}),    can be written as
 \be   \rho^{r}(\phi_f,\phi'_f;t) = \int D\phi_i   D\phi'_i  \,  \mathcal{T}[\phi_f,\phi'_f;\phi_i,\phi'_i;t] \,\rho_\phi(\phi_i,\phi'_i;0)\,, \ee
where the time evolution kernel is given by
\be \mathcal{T}[\phi_f,\phi_i;\phi'_f,\phi'_i;t] = {\int} \mathcal{D}\phi^+ \, \int \mathcal{D}\phi^- \, e^{i  \int d^4x \left[\mathcal{L}_0[\phi^+]-\mathcal{L}_0[\phi^-]\right]}\,e^{i\mathcal{I}[\phi^+;\phi^-]}\,, \ee
from which the non-equilibrium \emph{in-in effective action}   is identified as
\begin{equation}
    {S}_{eff}[\phi^+,\phi^-] = \int^t_0 dt' \int d^3 x \Big\{ \mathcal{L}_0[\phi^+]-\mathcal{L}_0[\phi^-] +\mathcal{I}[ \phi^+, \phi^-] \Big\} \,,\label{Leff}
\end{equation} where
$\mathcal{I}[\phi^+;\phi^-]$ is the \emph{influence action}\cite{feyver} obtained by tracing over the $\chi$ degrees of freedom,
\bea
    e^{i\mathcal{I}[\phi^+;\phi^-]} & = &  \int D\chi_i \,D\chi'_i D\chi_f  \int \mathcal{D}\chi^+ \int \mathcal{D}\chi^- \, e^{i  \int d^4x \left[\mathcal{L}[\chi^+]-  \sum_{a}g_a\,\phi^+_a\,\mathcal{O} [\chi^+] \right]}\nonumber \\  & \times &  e^{-i  \int d^4x \left[\mathcal{L}[\chi^-]-\sum_{b}g_b\,\phi^-_b \,\mathcal{O}[\chi^-] \right]} \,   \rho_{\chi}(\chi_i,\chi'_i;0)\,.
    \label{inffunc}
\eea

The path integral representations for both $\mathcal{T}[\phi_f,\phi_i;\phi'_f,\phi'_i;t]$ and $\mathcal{I}[\phi^+;\phi^-]$ feature the boundary conditions in (\ref{piforwardbc}-\ref{pibackbc}) except that we now set $\chi^\pm(\vec{x},t) = \chi_f(\vec{x})$ to trace over $\chi$ field.

In the above path integral defining the influence action  eqn. (\ref{inffunc}),  the   fields $ \phi^\pm_a(x) $ act as   \emph{external sources} (c-number) coupled to the   operator $\mathcal{O}[\chi]$. Therefore, it is straightforward to conclude that the right hand side of eqn. (\ref{inffunc}) is the path integral representation of the trace over the environmental fields coupled to external sources $\phi^\pm$, namely
\be e^{i\mathcal{I}[\phi^+;\phi^-]} = \mathrm{Tr}_{\chi} \Big[ \mathcal{U}(t;\phi^+)\,\rho_\chi(0)\,  \mathcal{U}^{-1}(t;\phi^-) \Big]\,, \label{trasa}\ee where   $\mathcal{U}(t;\phi^\pm)$ is the   time evolution operator in the $\chi$ Hilbert space in presence of \emph{external sources} $\phi^\pm$, i.e.  \be \mathcal{U}(t;\phi^+) = T\Big( e^{-i \int_0^t H_\chi[\phi^+(t')]dt'}\Big) ~~;~~
\mathcal{U}^{-1}(t;\phi^-) = \widetilde{T}\Big( e^{i \int_0^t H_\chi[\phi^-(t')]dt'}\Big)\,,\label{us} \ee
with \be H_\chi[\phi^\pm(t)] = H_{\chi}+ \int d^3x \, {\sum}_{a}g_a\,\phi^\pm_a(\vec{x},t)\mathcal{O}[\chi(\vx)]  \label{totiH} \ee   and $\widetilde{T}$ is the \emph{anti-time evolution operator} describing   evolution backwards in time, it is defined by $\widetilde{T}(A(t_1)B(t_2)) = A(t_1) B(t_2)\Theta(t_2-t_1)+B(t_2)A(t_1)\Theta(t_1-t_2)$.

The calculation of the influence action is facilitated by passing to the interaction picture for the Hamiltonian $H_\chi[\phi^\pm(t)]$, defining
\be  \mathcal{U}(t;\phi^\pm) = e^{-i H_{\chi}\,t} ~ \mathcal{U}_{ip}(t;\phi^\pm) \label{ipicture} \ee and the $e^{\pm i H_{\chi}\,t}$ cancel out in the trace in (\ref{trasa}), the interaction picture operators are now standard, and given by
 \bea  &&  \mathcal{U}_{ip}(t;\phi^+)   =   1-i\,   \int d^4 x' \sum_{a} g_a\, \phi^+_a(\vx',t')\,\mathcal{O}(\vx',t') \nonumber \\ & - & \frac{1}{2}\,  \int d^4 x_1\,  \int d^4 x_2 \sum_{a,b}g_a\,g_b\,\,T \Big(\phi^+_a(\vx_1,t_1)\,\mathcal{O}(\vx_1,t_1)\phi^+_b(\vx_2,t_2)\,\mathcal{O}(\vx_2,t_2)\Big)+\cdots \label{uplus}\eea
 \bea &&  \mathcal{U}^{-1}_{ip}(t;\phi^-)    =    1+i\,   \int d^4 x'\sum_{a} g_a\, \phi^-_a(\vx',t')\,\mathcal{O}(\vx',t') \nonumber \\ &  - & \frac{1}{2}\,  \int d^4 x_1\, \int d^4 x_2 \sum_{a,b}g_a\,g_b\,\,\widetilde{T}\Big(\phi^-_a(\vx_1,t_1)\,\mathcal{O}(\vx_1,t_1)\phi^-_b(\vx_2,t_2)\,\mathcal{O}(\vx_2,t_2)\Big)+\cdots \,,\label{umin} \eea where the composite operator  $\mathcal{O}(\vx,t) \equiv \mathcal{O}[\chi(\vx,t)]$ is in the Heisenberg picture of $H_{\chi}$.

Now the trace (\ref{trasa}) can be obtained systematically in perturbation theory in $g_a$, it is important to note that in taking the trace,  the operators with $\phi^+$  always appear in front of $\rho_{\chi}(0)$, whereas those with $\phi^-$  appear  behind, for example $\mathrm{Tr} \mathcal{O}^+(x_1) \rho_\chi(0) \mathcal{O}^-(x_2) = \mathrm{Tr} \mathcal{O}(x_2) \mathcal{O}(x_1) \rho_\chi(0) \equiv \langle  \mathcal{O}(x_2) \mathcal{O}(x_1) \rangle_{\chi}$, etc. Including all such terms up to quadratic order in $\phi^{\pm}$  we find
 \bea \mathcal{I}[\phi^+,\phi^-] & = &    -   \int d^4x \sum_{a} g_a \,\Big( \phi^+_a(x)-\phi^-_a(x)\Big)\,\langle \mathcal{O}(x)\rangle_\chi \nonumber + \\ & & \frac{i   }{2} \int d^4x_1 \int d^4x_2 \sum_{a,b} g_a\,g_b\,\Bigg\{ \phi^+_a(x_1) \,\phi^+_b(x_2)\,G^{++}(x_1-x_2)+ \phi^-_a(x_1)\,\phi^-_b(x_2)\,G^{--}(x_1-x_2) \nonumber \\
 & - & \phi^+_a(x_1)\,\phi^-_b(x_2)\,G^{+-}(x_1-x_2)- \phi^-_a(x_1)\,\phi^+_b(x_2)\,G^{-+}(x_1-x_2)\Bigg\}\,. \label{finF}\eea

 This result is confirmed by expanding the left hand side  of (\ref{trasa}) and comparing to the right hand side after symmetrization of the crossed terms of the form $\phi^+_a \phi^-_b$. In this expression  the \emph{connected} correlation functions in the initial density matrix of the $\chi$ fields, namely $\rho_\chi(0)$ are given by

  \begin{eqnarray}
&& G^{-+}(x_1-x_2) =   \langle
{\cal O}(x_1) {\cal O}(x_2)\rangle_\chi - \langle \mathcal{O}(x_1)\rangle_\chi  \langle \mathcal{O}(x_2)\rangle_\chi \equiv G^>(x_1-x_2) \,,\label{ggreat} \\&&  G^{+-}(x_1-x_2) =   \langle \mathcal{O}(x_2) \mathcal{O}(x_1)\rangle_\chi - \langle \mathcal{O}(x_2)\rangle_\chi  \langle \mathcal{O}(x_1)\rangle_\chi \equiv G^<(x_1-x_2)  \,,\label{lesser} \\&& G^{++}(x_1-x_2)
  =
{ G}^>(x_1-x_2)\Theta(t_1-t_2)+ {  G}^<(x_1-x_2)\Theta(t_2-t_1) \,,\label{timeordered} \\&& G^{--}(x_1-x_2)
  =
{ G}^>(x_1-x_2)\Theta(t_2-t_1)+ {  G}^<(x_1-x_2)\Theta(t_1-t_2)\,,\label{antitimeordered}
\end{eqnarray}
 in terms of fields in the Heisenberg picture of $H_\chi$, where
\be \langle (\cdots) \rangle_{\chi} = \mathrm{Tr}_\chi(\cdots)\rho_\chi(0)\,. \label{expec}\ee

We highlight  that the correlation functions $G^{>},G^{<}$ are \emph{exact}, namely to \emph{all orders} in the couplings of the environmental fields $\chi$ that enter in $\mathcal{O}$ to all other fields to which it couples except the fields $\phi_{1,2}$.

We will assume that the expectation value of the  composite operator $\mathcal{O}[\chi]$ has been subtracted if necessary so that $\langle \mathcal{O}[\chi]\rangle =0$, therefore the first term in (\ref{finF}) vanishes along with the second terms on the right hand sides of eqns. (\ref{ggreat},\ref{lesser}).

Since the operator  $\mathcal{O}$ is hermitian, it follows from the definitions above that
\be G^< (x_1-x_2) = G^> (x_2-x_1)\,. \label{idt}\ee

The influence action (\ref{finF}) becomes simpler by writing it  solely in terms of the two correlation functions $G^\lessgtr$, this is achieved by implementing the following steps:

 \begin{itemize}
\item{In the term with $\phi^+_a(x_1)\phi^+_b(x_2)$: in the contribution $G^< (x_1-x_2)\Theta(t_2-t_1)$ (see eqn. (\ref{timeordered})) relabel $x_1 \leftrightarrow x_2$, $a \leftrightarrow b$ and use the property (\ref{idt}).  }
\item{ In the term with $\phi^-_a(x_1)\phi^-_b(x_2)$: in the contribution $G^> (x_1-x_2)\Theta(t_2-t_1)$ (see eqn. (\ref{antitimeordered})) relabel $x_1 \leftrightarrow x_2$, $a \leftrightarrow b$ and use the property (\ref{idt}). }
    \item{ In the term with $\phi^+_a(x_1)\phi^-_b(x_2)$: multiply $G^< (x_1-x_2)$ by $\Theta(t_1-t_2)+\Theta(t_2-t_1)=1$ and in the term with $\Theta(t_2-t_1)$ relabel $ x_1 \leftrightarrow x_2$, $a \leftrightarrow b$ and use the property (\ref{idt}). }
 \item{ In the term with $\phi^-_a(x_1)\phi^+_b(x_2)$: multiply $G^> (x_1-x_2)$ by $\Theta(t_1-t_2)+\Theta(t_2-t_1)=1$ and in the term with $\Theta(t_2-t_1)$ relabel $x_1 \leftrightarrow x_2$, $a \leftrightarrow b$   and use the property (\ref{idt}). }
\end{itemize}
We find
\bea  &&  \mathcal{I}[\phi^+, \phi^-]    = i\, \sum_{a,b}  g_a\,g_b\int d^4x_1 d^4x_2  \,\Bigg\{ \phi^+_a(\vx_1,t_1)\phi^+_b(\vx_2,t_2)\,G^> (x_1-x_2) +   \phi^-_a(\vx_1,t_1)\phi^-_b(\vx_2,t_2)\,G^<(x_1-x_2) \nonumber\\
  &- & \phi^+_a(\vx_1,t_1)\phi^-_b(\vx_2,t_2)\,G^<(x_1-x_2)  -   \phi^-_a(\vx_1,t_1)\phi^+_b(\vx_2,t_2)\,G^>(x_1-x_2)\Bigg\}\Theta(t_1-t_2) \label{Funravel}\eea where $G^{\lessgtr}$ are given by eqns. (\ref{ggreat},\ref{lesser}).  This is the general form of the influence function up to second order in the  couplings of the fields $\phi_{1,2}$ to the environmental fields $\chi$  but \emph{to all orders} in the couplings of the environmental fields  to \emph{any} other field except $\phi_{1,2}$. Notice that $\mathcal{I}[\phi^+, \phi^-]\Big|_{\phi^+=\phi^-}=0$ consistently with its definition given by eqn. (\ref{trasa}).
  The contributions to the influence action $\mathcal{I}[\phi^+, \phi^-]$ can be interpreted in terms of self-energy type diagrams shown in fig. (\ref{fig:influence}).
       \begin{figure}[ht]
\includegraphics[height=2.5in,width=2.5in,keepaspectratio=true]{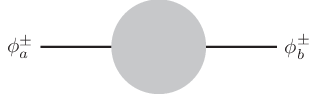}
\caption{Contributions to the influence functional $\mathcal{I}[\phi^+, \phi^-]$. The black circle denotes the correlation functions $G^{\lessgtr}$ to all orders in the couplings to degrees of freedom other than the fields $\phi_{a,b}$.}
\label{fig:influence}
\end{figure}
For the case of pion-axion mixing, the type of self-energy diagrams that determine the influence action is shown in fig. (\ref{fig:pionaxion}), and explicitly shows the origin of axion-pion mixing. We emphasize that whereas fig. (\ref{fig:pionaxion}) displays a one-loop  self-energy as an illustrative example, the results are valid to all orders in the couplings of the electromagnetic field to any other charged field within or beyond the standard model. This is explicit in the exact Lehmann-representation of the correlation functions analyzed in appendix (\ref{app:correlations}).  In fig. (\ref{fig:pionaxion}) the photon propagators can be  dressed by loops of the charged fields, for example $e^+e^-$ loops, and similarly for quark-antiquark loops, describing processes such as    $\pi^0 \gamma \leftrightarrow e^+ e^- \leftrightarrow   \gamma\, a$, etc. These radiative corrections to the photon propagators result in a thermal mass for the photon\cite{htl}, all of these processes are included in the exact spectral representation of the Green's functions.
         \begin{figure}[ht]
\includegraphics[height=2.5in,width=2.5in,keepaspectratio=true]{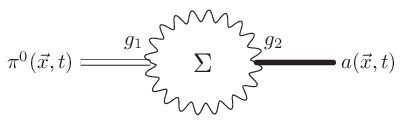}
\caption{Contributions to the influence functional for the pion-axion mixing case. The     self-energy diagram corresponds to the processes $\pi^0 \leftrightarrow 2\gamma^* \leftrightarrow a$   and the inverse process,  with $\gamma^*$ a photon in the medium. These processes induce axion-pion mixing. The photon propagators can be dressed by loops of charged particles. The branch labels $\pm$ are suppressed for clarity. }
\label{fig:pionaxion}
\end{figure}

 We can obtain expectation values and correlation functions of the meson fields by including sources $J^{\pm}_a(x)$ with $\mathcal{L}_0(\phi^\pm)\rightarrow \mathcal{L}_0(\phi^\pm)\pm{\sum}_a J^{\pm}_a(x)\phi^\pm_a(x)$ and defining the generating functional
  \be \mathcal{Z}[J^+,J^-] = \mathrm{Tr}\,\rho^r(J^+,J^-;t) =  \int D\phi_f D\phi_i   D\phi'_i  \,  {\int} \mathcal{D}\phi^+ \, \int \mathcal{D}\phi^- \, e^{i S_{eff}[\phi^+,J^+;\phi^-,J^-;t]} \,\rho_\phi(\phi_i,\phi'_i;0) \label{Zofj}\ee with the boundary conditions
  \bea &  &   \phi^+_a(\vec{x},t=0)=\phi_{i,a}(\vec{x})~;~
 \phi^+_a(\vec{x},t)  =   \phi_{f,a}(\vec{x}) \nonumber \\
&  &   \phi^-_a(\vec{x},t=0)=\phi'_{i,a}(\vec{x})~;~
 \phi^-_a(\vec{x},t)  =   \phi_{f,a}(\vec{x}) \,.\label{bctraza}\eea Expectation values or correlation functions of $\phi^{\pm}$ in the reduced density matrix are obtained as usual with variational derivatives with respect to the sources $J^\pm$.

 The effective action (\ref{Leff}) may be written in a manner more suitable to exhibit the equations of motion by introducing the Keldysh\cite{keldysh} variables
\be \Phi_a(\vx,t')= \frac{1}{2}\big( \phi^+_a(\vx,t')+\phi^-_a(\vx,t')\big)~~;~~\mathcal{R}_a(\vx,t')= \big( \phi_a^+(\vx,t')-\phi^-_a(\vx,t')\big)\label{kelvars} \,. \ee

The boundary conditions on the $a^\pm$ path integrals given by
(\ref{bctraza}) translate into the following boundary conditions on
the center of mass and relative variables
\begin{align}
    \Phi_a(\vec x,t=0)= \Phi_{a,i} \; \; &; \; \; \mathcal{R}_a(\vec x,t=0)=\R_{a,i} \,, \label{bcwig} \\
    \Phi_a(\vec{x},t ) = \Phi_{a,f}(\vec{x}) \; \; &; \; \; \mathcal{R}_a(\vec x,t  )=0 \,. \label{Rfin}
\end{align}

 In terms of the center of mass and relative field variables, the effective action (\ref{Leff})  with the influence functional (\ref{finF}) in terms of spatial Fourier transform becomes
\bea
   && iS_{eff}[\Phi,\R] =
      - i \int d^3x \,\sum_{a}\,\mathcal{R}_{a,i}(x) \dot{\Phi}_a(x,t=0) \nonumber \\
    & & - i \int_0^t dt_1\, \sum_{\vec{k}}   \mathcal{R}_a(-\vec{k},t_1)  \left\{ \ddot{\Phi}_{a}(\vec{k},t_1) + \big(\omega^2_a(k)\delta_{ab} +  \Delta m^2_{ab}\big) \Phi_{b}(\vk,t_1)  -  \Phi_a (\vec{k},t_1) \J_a(-\vec{k},t_1) \right\} \nonumber \\
    && - \int_0^t dt_1 \int_0^t dt_2 \sum_{\vk}\,\left\{ \frac{1}{2} \mathcal{R}_a(-\vec{k}, t_1) \mathcal{N}_{ab}(\vec{k};t_1-t_2) \R_b(\vec{k},t_2) + i\,\mathcal{R}_a(-\vec{k},t_1)   \Sigma^R_{ab}(\vec{k};t_1 - t_2) \Phi_b(\vec{k}; t_2) \right\}\,.\nonumber \\
    \label{efflanwig}
\eea where repeated indices are summed over, and $\omega^2_{a}(k) = k^2+m^2_a$. To obtain the above form,  we integrated by parts,  defined $ \J_a(x)= (J^+_a(x)-J^-_a(x))$, and  kept only the sources conjugate to $\Phi_a$ because we are interested in expectation values and correlation functions of this variable only as discussed in detail below.

The kernels in the above effective Lagrangian are given by (see
eqns. (\ref{ggreat}-\ref{antitimeordered}))
\begin{eqnarray}
\mathcal{N}_{ab}(k,t-t') & = & \mathbb{C}_{ab}\,
\frac{1}{2} \left[   G^>(k;t-t')+{ G}^<(k;t-t') \right] \label{kernelkappa} \\
i\Sigma^{R}_{ab}(k,t-t') & = &   \mathbb{C}_{ab}\, \left[{ G}^>(k;t-t')-{
G}^<(k;t-t') \right]\Theta(t-t') \equiv
i\Sigma_{ab}(k,t-t')\Theta(t-t') \label{kernelsigma}
\end{eqnarray} where $G^{<,>}(k;t-t')$ are the spatial Fourier transforms of the correlation functions in (\ref{ggreat}-\ref{antitimeordered}) and
\be \mathbb{C}_{ab} =  \left(
                         \begin{array}{cc}
                           g^2_1 & g_1g_2 \\
                           g_1g_2 & g^2_2 \\
                         \end{array}
                       \right)\,.
\label{coupmtx}\ee
 In the exponential of the effective action $e^{iS_{eff}}$, the quadratic term in the relative variables $\R_a$ is written as a functional integral over a noise variable $\xi_a$ as follows,

\bea
    &&\exp\left\{ -  \frac{1}{2}\,\int dt_1 \int dt_2   \,\mathcal{R}_a(-\vec{k}; t_1) \mathcal{N}_{ab}(\vec{k};t_1 - t_2) \mathcal{R}_b(\vec{k};t_2) \right\} =     \nonumber \\
    &&   A \int \mathcal{D}\xi_a \, \exp\left\{ - \frac{1}{2} \int dt_1 \int dt_2 \, \xi_a(-\vec{k};t_1) \mathcal{N}^{-1}_{ab}(\vec{k};t_1-t_2) \xi_b(\vec{k};t_2) + i \int dt \xi_a(-\vec{k};t) \R_a(\vec{k};t) \right\}\nonumber\\
    \label{nois}
\eea
where $A$ is a normalization factor.

We seek to obtain the equations of motion as an initial value problem rather than a boundary value problem. This is achieved by writing  the initial density matrix in terms of the initial center of mass and relative variables $\Phi_{a,i},\R_{a,i}$ as
\be \rho_\phi(\phi_{a;i},\phi'_{a;i};0) \equiv \rho_\phi(\Phi_{a,i}+\frac{\R_{a,i}}{2},\Phi_{a,i}-\frac{\R_{a,i}}{2};0)\,, \label{rhoavars}\ee and by  introducing the functional Wigner transform\cite{zubairy}
\be W[\Phi_{a,i},\Pi_{a,i}] = \int D\R_i \, e^{-i \int d^3 x \Pi_{a,i}(\vx) \,\R_i(\vx)}\,\rho_\phi(\Phi_{a,i}+\frac{\R_{a,i}}{2},\Phi_{a,i}-\frac{\R_{a,i}}{2};0) \,,\label{wigner}\ee which allows us to write (up to a normalization factor)
\be \rho_\phi(\Phi_{a,i}+\frac{\R_{a,i}}{2},\Phi_{a,i}-\frac{\R_{a,i}}{2};0)    = \int D\Pi_{a,i} \, e^{i \int d^3 x  \Pi_{a,i}(\vx) \,\R_{a,i}(\vx)}\, W[\Phi_{a,i},\Pi_{a,i}]
\,. \label{invwig}\ee

The Wigner transform is the ``closest'' to a (semi) classical phase space distribution function\cite{zubairy}.

Gathering these results together, we now write the generating functional (\ref{Zofj}) in terms of the Keldysh variables (\ref{kelvars}), with the effective action in these variables given by eqn. (\ref{efflanwig}),
 implementing the Wigner transform (\ref{invwig}) and using the representation (\ref{nois}) we obtain
 \begin{align}
    & \mathcal{Z}[\mathcal{J}] =
        \int D\Phi_f \int D\R_i\, D\Phi_i\, D\Pi_i \int D\Phi\, D\R\, D\xi\; W[\Phi_i,\Pi_i] \times P[\xi] \times \exp\left\{ i \int dt \sum_{\vec{k}} \Phi_a(\vec{k};t) \J_a(-\vec{k};t) \right\}\,\times
    \nonumber \\
    &  \exp\left\{ -i \int dt \sum_{\vec{k}}   \R_a(-\vec{k};t) \Bigg( \ddot{\Phi}_a(\vec{k};t) + \big(\omega^2_a(k)\delta_{ab}+\Delta m^2_{ab}\big)  \Phi_b(\vec{k};t) + \int_0^t \Sigma_{ab}(\vec{k};t-t') \Phi_b(\vec{k},t') dt' - \xi_a(\vec{k};t) \Bigg)   \right\}
    \nonumber \\
    & \times \exp\left\{ i \sum_{\vec{k}} \R_{a,i}(-\vec{k}) \left( \Pi_{a,i}(\vec{k}) - \dot{\Phi}_{a,i}(\vec{k}) \right)  \right\}\,,
    \label{Zetafinal}
\end{align}
 where repeated field indices are summed over, and the noise probability distribution function $P[\xi]$ is given by


\begin{equation}
    P[\xi] = A\, \prod_{\vec{k}} \exp\left\{ -\frac{1}{2} \int  dt_1 \int  dt_2\, \xi_a(-\vec k;t_1)\,{\mathcal{N} }^{-1}_{ab}(k;t_1-t_2)\,\xi_b(\vec k;t_2) \right\}\,.
    \label{noispdf}
\end{equation}
The generating functional $\mathcal{Z}[\J]$ is the final form of the time evolved reduced density matrix   after tracing over the bath degrees of freedom. Variational derivatives with respect to the source $\J$ yield  the correlation functions of the Keldysh center of mass variables $\Phi$\cite{beilok}.

Carrying out the functional integrals over $\R_i(\vec{k})$ and $\R_{\vec{k}}(t)$ yields a more clear form, namely
\begin{align}
    \mathcal{Z}[\mathcal{J}] \propto
   & \int D\Phi_{a,f} \int D\Phi_{a,i}\, D\Pi_{a,i} \int D\Phi_a\, D\xi_a\; W[\Phi_i,\Pi_i] \times P[\xi] \times \exp\left\{ i \int dt \sum_{a,\vec{k}} \Phi_a(\vec{k};t) \J_a(-\vec{k};t) \right\}
    \nonumber \\
    & \times \prod_{\vec{k}} \delta\left[ \ddot{\Phi}_a(\vec{k};t) + (\omega^2_a({k})\delta_{ab}+\Delta m^2_{ab}) \, \Phi_b(\vec{k};t) + \int_0^t \Sigma_{ab}(\vec{k};t-t') \Phi_b(\vec{k};t') dt' - \xi_a(\vec{k};t) \right]\nonumber \\
      & \times \prod_{a,\vec{k}} \delta\left[ \Pi_{a,i}(\vec{k}) - \dot{\Phi}_{a,i}(\vec{k}) \right]\,.
    \label{Zetadelta}
\end{align}
The functional delta functions  determine   the field configurations that contribute  to the generating functional $\mathcal{Z}[\mathcal{J}]$:
 \begin{itemize}
 \item The equation of motion of $\Phi_a(\vec{k};t)$ is a   \emph{stochastic} Langevin equation\cite{beilok,shuyang}, namely
    \begin{equation}
        \ddot{\Phi}_a(\vec{k};t) + \omega^2_a( {k})\,\Phi_a(\vk,t)+\Delta m^2_{ab}\,\Phi_b(\vk,t)\,   + \int_0^t \Sigma_{ab}(\vec{k};t-t') \Phi_b(\vec{k};t') dt' = \xi_a(\vec{k};t)\,.\label{langevin}
    \end{equation}
    Note that this equation of motion involves the \emph{retarded} self-energy, thereby defining a causal initial value problem, this is a distinct consequence of the in-in formulation of time evolution.

    \item The initial conditions of $\Phi_a(\vec{k})$ satisfy
    \begin{equation}
        \Phi_a(\vec{k};t=0) = \Phi_{a,i}(\vec{k}) \qquad;\qquad \dot{\Phi}_a(\vec{k};t=0) = \Pi_{a,i}(\vec{k})\,, \label{inicons}
    \end{equation}
    where $\Phi_{a,i}(\vec{k}),\Pi_{a,i}(\vec{k})$ are distributed according to the probability distribution function $W[\Phi_{a,i},\Pi_{a,i}]$ in turn determined by the  the initial density matrix. This is one of the manifestations of stochasticity. We introduce the notation $\overline{(\cdots)}$ to denote averaging over the initial conditions (\ref{inicons})  with the distribution function $W[\Phi_{a,i},\Pi_{a,i}]$.

    \item The expectation value and correlations of the stochastic noise $\xi_a(\vec{k};t)$ are determined by the Gaussian probability distribution $P[\xi]$, yielding
    \be
        \langle\langle \xi_a(\vk,t)  \rangle \rangle =0
        ~~;~~
        \langle\langle \xi_a(\vk;t) \xi_b(\vk';t')  \rangle \rangle = \mathcal{N}_{ab}(k;t-t')\,\delta_{\vk,-\vk'} \,, \label{noiscors}
    \ee
    where $\langle\langle \cdots \rangle\rangle$ means averaging weighted by $P[\xi ]$. Since $P[\xi ]$ is Gaussian, higher order correlation functions are obtained by implementing Wick's theorem. This averaging is the second manifestation of stochasticity. It is important to highlight that Gaussianity of the noise does not imply a free field theory, the kernel $\mathcal{N}_{ab}(k;t-t')$ is in principle to all orders in the coupling of photons to charged fields within or beyond the standard model, and is at least of one-loop order as displayed in fig.(\ref{fig:pionaxion}) and explained in detail in appendix (\ref{app:correlations}).
\end{itemize}

The solutions of the Langevin equation (\ref{langevin}) $\Phi_a[\vk;t;\xi;\Phi_{a,i};\Pi_{a,i}]$ are \emph{functionals} of the stochastic noise  variables $\xi_a$ and the initial conditions,  therefore  correlation functions of the original field variables $\phi_a$  in the reduced density matrix correspond to  averaging the products of the solutions over both the initial conditions with the Wigner distribution function, and the noise with the probability distribution function $P[\xi]$. We denote such averages  by $\overline{\langle\langle \big( \cdots \big) \rangle\rangle}$ with $\big( \cdots \big)$   any functional of the initial conditions (\ref{inicons}) and $\xi_a$. These stochastic averages  yield  the expectation values and correlation functions of functionals of $\Phi_a$ obtained from variational derivatives with respect to $\mathcal{J}_a$.

We seek to obtain correlation functions of $\phi_a$ in the reduced density matrix, therefore we must relate these to the averages of the center of mass Keldysh fields $\Phi_a$. To establish this relation, we begin with the path integral representations for the forward and backward time evolution operators (\ref{evolrhot}, \ref{piforward},\ref{piback}) which show that $\phi^+_a$ are associated with $U(t)$ and $\phi^-_a$ with $U^{-1}(t)$, hence it follows that  inside the path integral operators    in the forward, backward and mixed forward-backward branches, are given by

 \bea && \mathrm{Tr}\phi^+_a(\vk;t)\phi^+_b(\vk';t')\,\rho(0) \equiv \mathrm{Tr}\,T\big(\phi_a (\vk;t)\phi_b (\vk';t'))\, \rho(0) \nonumber \\
&& \mathrm{Tr} \phi^-_a(\vk;t)\phi^-_b(\vk';t')\,\rho(0) \equiv \mathrm{Tr} \rho(0)\,\widetilde{T}\big( \phi_a(\vk;t)\phi_b(\vk';t')\big) \nonumber \\
 && \mathrm{Tr} \phi^+_a(\vk;t)\phi^-_b(\vk';t')\,\rho(0) \equiv  \mathrm{Tr} \phi_a(\vk;t)\, \rho(0)\, \phi_b(\vk';t') = \mathrm{Tr}\,\phi_b(\vk';t') \, \phi_a(\vk;t)\, \rho(0)  \,,\label{dict}\eea where $T,\widetilde{T}$ are the time ordering and anti-time ordering symbols respectively. From the cyclic property of the trace, it follows that the expectation value of the $\phi$ fields in the total density matrix is
\be  \langle \phi_a(\vx,t) \rangle   =   \mathrm{Tr}\phi^+_a(\vx,t)\,\hat{\rho}(0)=   \mathrm{Tr}  {\rho}(0)\,\phi^-_a(\vx,t)=  \mathrm{Tr} \Phi_a(\vx,t)\, {\rho}(0) = \overline{\langle \langle \Phi_a(\vx,t) \rangle \rangle} \,,\label{averageA} \ee
whereas
\be \mathrm{Tr} \R_a(\vx,t)\,\hat{\rho}(0) =0 \label{aveR}\,. \ee Furthermore, using the relations (\ref{dict}) it is straightforward to confirm that
\be \mathrm{Tr}\,\Phi_a(\vx,t)\Phi_b(\vx',t')\,\rho(0) \equiv \frac{1}{2}\,\mathrm{Tr}\Big( \phi_a(\vx,t)\phi_b(\vx',t')+\phi_b(\vx',t')\phi_a(\vx,t)\Big)\,\rho(0)\,.\label{ide} \ee

Upon obtaining the functional solutions of eqn. (\ref{langevin}) our objective is to obtain the connected equal time correlation functions
\be \langle \phi_a(t) \phi_b(t) \rangle_c= \mathrm{Tr} \rho(0) \phi_a(t) \phi_b(t) -  \mathrm{Tr} \rho(0) \phi_a(t) \, \mathrm{Tr} \rho(0)   \phi_b(t)\,, \label{conncorr} \ee which provides a measure of
\emph{coherence}, and the population of each mode of wavevector $k$, namely
\be n_a(k;t) = \frac{1}{2\omega_a(k)}\,\mathrm{Tr}\rho(0)\, \Big[\dot{\phi}_a(\vk;t)\dot{\phi}_a(-\vk;t)+\omega^2_a(k) \, {\phi}_a(\vk;t) {\phi}_a(-\vk;t)\Big]  - \frac {1}{2} ~~~~(\mathrm{no~sum~over~a}) \,.\label{numbs}\ee

The fields $\phi_a(t);\phi_b(t)$   commute at equal times, therefore with the definition of the Keldysh center of mass field variables $\Phi_a$ (\ref{kelvars}) and the relations (\ref{dict},\ref{averageA},\ref{aveR}) we find that the equal time connected correlation function (\ref{conncorr}) is given by
\be \langle \phi_a(t) \phi_b(t) \rangle_c= \overline{\langle\langle \Phi_a(t) \Phi_b(t) \rangle \rangle}- \overline{\langle\langle \Phi_a(t)   \rangle \rangle} ~~ \overline{\langle\langle   \Phi_b(t) \rangle \rangle}\,.\label{conncorrfin}\ee In analogy with the density matrix of two level systems\cite{zubairy} or qubits, we refer to the off-diagonal connected correlation function as coherences.

To obtain the population per mode (\ref{numbs}) we   introduce the correlation functions (no sum over the label a)
\be \mathcal{C}^>_a(k;t,t') = \mathrm{Tr} \phi_a(\vk;t) \phi_a(-\vk;t')  {\rho}(0)~~;~~ \mathcal{C}^<_a(k;t,t') = \mathrm{Tr} \phi_a(-\vk;t') \phi_a(\vk;t) {\rho}(0)\,, \label{gis}\ee  the populations per mode of wavevector $\vk$ (\ref{numbs}) become
\be n_a(k;t) = \frac{1}{4\,\omega_a(k)}\,  \Bigg(\frac{\partial}{\partial t}\frac{\partial}{\partial t'} +\omega^2_a(k)  \Bigg)\Bigg[\mathcal{C}^>_a(k;t,t')+\mathcal{C}^<_a(k;t,t') \Bigg]_{t=t'}-\frac{1}{2} \,. \label{number} \ee   Using the identity (\ref{ide})  it is straightforward to show that this symmetrized product yields
\bea n_a(k;t)  & = &   \frac{1}{2\omega_a(k)}\,\Bigg\{\overline{\langle \langle \dot{\Phi}_a(\vk;t)\dot{\Phi}_a(-\vk;t)\rangle \rangle}+ \omega^2_a(k) \,\overline{\langle \langle {\Phi}_a(\vk;t) {\Phi}_a(-\vk;t)  \rangle \rangle}\Bigg\}-\frac{1}{2} \,,\label{eneA} \eea and the energy per mode
\be \mathcal{E}_a(k;t) \equiv   \Big( n_a(k;t)+\frac{1}{2}\Big)\,\omega_a(k) \,.\label{energia}\ee
The corollary of this analysis is that we can obtain the connected correlation functions   and the populations of the fields $\phi_{1,2}$ by obtaining the solutions of the Langevin equation of motion (\ref{langevin}) with initial conditions (\ref{inicons}) and taking the averages over the initial conditions and noise described above.  The stochastic equation of motion (\ref{langevin}) with initial conditions (\ref{inicons}) defines an initial value problem whose  solution   is obtained by Laplace transform. The Laplace transforms are given by
\bea \widetilde{\Phi}_a(\vk;s) & = &  \int^\infty_0 e^{-st}\,\Phi_a(\vk;t)\,dt \,,\label{laplaA}    \\ \widetilde{\xi}_a(\vk;s) & = &  \int^\infty_0 e^{-st}\,\xi_a(\vk;t)\,dt \,,\label{laplachi} \\ \widetilde{\Sigma}_{ab}(\vk;s) & = &  \int^\infty_0 e^{-st}\,\Sigma_{ab}(\vk;t)\,dt  = -\frac{\mathbb{C}_{ab}}{2\pi} \,\int^{\infty}_{-\infty} \frac{\rho(k_0,k)}{k_0-is} dk_0 \,,\label{laplasigma}
\eea where in (\ref{laplasigma}) we used the representation (\ref{sigmadis}). The Laplace transform of the Langevin eqn. (\ref{langevin}) becomes
\be \mathbb{G}^{-1}_{ab}(k,s)\,\widetilde{\Phi}_b(\vk;s)= \Pi_{a,i}(\vk)+ s\,\Phi_{a,i}(\vk)+ \widetilde{\xi}_a(\vk;s)\,,\label{laplalang}\ee
where
\be \mathbb{G}^{-1}_{ab}(k,s) = (s^2+\omega^2_a)\delta_{ab}+ \Delta m^2_{ab}+\widetilde{\Sigma}_{ab}(\vk;s)\,. \label{inverG} \ee

The solution in real time is obtained by inverse Laplace transform, it is given by
\be \Phi_a(\vk;t) = \Phi^h_a(\vk;t) +  \Phi^{\xi}_a(\vk;t)\,, \label{Asplits} \ee  where $\Phi^h_a(\vk;t); \Phi^{\xi}_a(\vk;t)$ are the homogeneous   and inhomogeneous solutions respectively, namely

\bea \Phi^h_a(\vk;t)  & =  & \dot{\mathcal{G}}_{ab}(k;t)\Phi_{b,i}(\vk)  +  \,\mathcal{G}_{ab}(k;t)\Pi_{b,i}\,, \label{homog}  \\  \Phi^{\xi}_a(\vk;t) & =  & \int^t_0 \mathcal{G}_{ab}(k;t-t')\,\xi_b(\vk;t')\,dt' \,,\label{realtisol} \eea  and the Green's function is given by
\be \mathcal{G}_{ab}(k;t) = \frac{1}{2\pi i} \int_{\mathcal{C}}   e^{st} \, \mathbb{G}_{ab}(k,s)\, ds \,, \label{goftsol}\ee   $\mathcal{C}$ denotes the Bromwich contour parallel to the imaginary axis   and to the right of all the singularities   of $\mathbb{G}_{ab}(k,s)$ in the complex s-plane and closing along a large semicircle at infinity with $Re(s)<0$. These singularities correspond to poles and multiparticle branch cuts with $Re(s)<0$, thus the contour runs parallel to the imaginary axis $s= i(\nu -i \epsilon)$, with $-\infty \leq \nu \leq \infty$ and $\epsilon \rightarrow 0^+$. Therefore,
\be \mathcal{G}_{ab}(k;t) =   \int^{\infty}_{-\infty}  \mathbb{G}_{ab}(k,s=i(\nu-i\epsilon))\, {e^{i\nu t}} \,\frac{d\nu}{2\pi}\,. \label{Goftfin}\ee We obtain the Green's function $\mathbb{G}_{ab}(k,s)$   in appendix (\ref{app:green}), it is given by eqn. (\ref{greenfin}) for the general case.

\vspace{1mm}

  The general form of the Green's function has been found in appendix (\ref{app:green}), it is given by eqns. (\ref{greenfin},\ref{capR}) with
\bea && M^2(s,k) = s^2+ k^2+ \frac{1}{2}\Big[m^2_1+m^2_2+ \Delta m^2_{11}+\Delta m^2_{22}+(g^2_1+g^2_2)\,\sigma(s,k) \Big] \label{M2}\\
&& \mathcal{D}(s,k) = \Bigg[\Big(m^2_1-m^2_2+ \Delta m^2_{11}-\Delta m^2_{22}+(g^2_1-g^2_2)\,\sigma(s,k)\Big)^2+4\Big(\Delta m^2_{12}+g_1g_2\,\sigma(s,k)\Big)^2  \Bigg]^{1/2} \nonumber \\ \label{capD}\\&& \alpha(s,k) = \frac{1}{\mathcal{D}}\,\Bigg[ m^2_1-m^2_2+ \Delta m^2_{11}-\Delta m^2_{22}+(g^2_1-g^2_2)\,\sigma(s,k)  \Bigg]\,,\label{alfi}\\ && \beta(s,k) = \gamma(s,k)= \frac{2}{\mathcal{D}}\Big[\Delta m^2_{12}+g_1g_2\,\sigma(s,k)\Big]\,, \label{beti} \eea with
\be \sigma(s,k) = -  \,\int^{\infty}_{-\infty} \frac{\rho(k_0,k)}{k_0-is} \frac{dk_0}{2\pi}\,.\label{sigma}  \ee  The counterterms $\Delta m^2_{ab}$ will be chosen to cancel the divergences in the self-energy in the effective theory with a cutoff $\Lambda$. The Green's function (\ref{Goftfin}) requires the analytic continuation $s\rightarrow i(\nu-i\epsilon)$ in the self energy, yielding
\be \sigma(\nu,k) = -\int^\infty_{-\infty} \mathcal{P}\Big[\frac{\rho(k_0,k)}{k_0-\nu} \Big]\,\frac{dk_0}{2\pi}\,+ i \, \frac{\rho(\nu,k)}{2}\,, \label{sigreim}\ee where we used the property $\rho(k_0,k) = -\rho(-k_0,k)$ (see appendix \ref{app:correlations}), as a consequence of which the real (R) and imaginary (I) parts of $\sigma(\nu,k)$ obey the property
\be \sigma_R(\nu,k) = \sigma_R(-\nu,k)~~;~~ \sigma_I(\nu,k)= -\sigma_I(-\nu,k)\,.\label{sigprope}\ee

\vspace{1mm}

\subsection{Renormalization aspects: kinetic mixing.}\label{subsec:renormalization}

Before we move on to the dynamical evolution, we now address several subtle aspects associated with renormalization. The couplings $g_{1,2}$ each feature mass dimension $(mass)^{-1}$, therefore the effective field theory is non-renormalizable, since the self-energy must be of mass dimension $(mass)^2$,     it follows that $\sigma$ must feature dimension $(mass)^4$. Although the full spectral density $\rho(k_0,k)$ is not available to all orders in perturbation theory, we can illustrate the main aspects of renormalization by considering the one photon loop contribution to the self-energy   obtained in ref.\cite{shuyang} and  given by eqns. (\ref{rhozero},\ref{rhoT})  in appendix (\ref{app:correlations}).

The zero temperature contribution to the spectral density (\ref{rhozero}) yields an ultraviolet divergent result for $\sigma_R(\nu,k)$, calculating the integral in (\ref{sigreim}) with an upper cutoff $\Lambda$ we find for the zero temperature part
\be  \sigma^{(0)}_R(\nu,k) = - \frac{1}{128\pi^2}  \Bigg\{    \Lambda^4  + 4\,K^2 \Lambda^2    + 2\,(K^2)^2\, \ln\Big[\frac{\Lambda^2\,e^{3/2}}{|K^2|}\Big]\Bigg\}~~;~~ K^2 = \nu^2-k^2 \,, \label{sigcero}
\ee   the finite temperature contribution does not yield ultraviolet divergences because of the thermal suppression associated with the distribution functions.  Whereas the first term $\propto \Lambda^4$ may be absorbed by mass counterterms, the terms $K^2\Lambda^2$ and $(K^2)^2\,\ln\Lambda$ imply the necessity for  off-diagonal four momentum dependent counterterms of the form
\be \partial_{\mu}\pi^0 \partial^{\mu} a ~~;~~ (\partial_\mu \partial^\mu\,\pi^0)(\partial_\nu \partial^\nu\,a)\,.\label{kinmix}\ee

This observation suggests  that the most general effective field theory mixing axions and neutral pions via a common intermediate state of photons \emph{must} include kinetic mixing terms of the form (\ref{kinmix}). This analysis leads us to conclude that mass counterterms of the form $\Delta m^2_{ab}$ are insufficient to absorb the ultraviolet divergences in the effective field theory and that other counterterms associated with the  kinetic mixing (\ref{kinmix}) must be included.

In the analysis that follows, we will \emph{assume} that such counterterms have been introduced to cancel all of the zero temperature divergences in the self-energies and $m^2_a$ are the renormalized masses. Hence we now set $\Delta m^2_{ab}=0$ in eqns. (\ref{M2}-\ref{beti}) along with the counterterms from kinetic mixing and keep solely the finite temperature (ultraviolet finite)
contributions to $\sigma$. Furthermore, we will also assume that the renormalized kinetic mixing terms all feature vanishing coefficients (after absorbing the ultraviolet divergent contributions from self-energies). Of course this is a special choice of the effective field theory, which we accept here without further
elaboration, however the  consequences of such kinetic mixing merits further study, which is  beyond the scope of this article.

\subsection{Dynamics:}\label{subsec:dynamics}
In order to obtain the solutions of the Langevin eqn. (\ref{Asplits}) from which we can extract the correlation functions and populations, it remains to obtain the Green's function (\ref{Goftfin}). Using the results of appendix  (\ref{app:green}) and after renormalization as described above, we obtain
\be \mathcal{G}_{ab}(k;t) =   \int^{\infty}_{-\infty}\Bigg\{\frac{\mathbb{P}_-(\nu,k)}{M^2(\nu,k) - \frac{\D(\nu,k)}{2} } +  \frac{\mathbb{P}_+(\nu,k)}{M^2(\nu,k) + \frac{\D(\nu,k)}{2} } \Bigg\}\,e^{i\nu t} \, \frac{d\nu}{2\pi} \,,\label{goft2}      \ee where
\be \mathbb{P}_{\pm}(\nu,k) = \frac{1}{2}  \left(
                                                                           \begin{array}{cc}
                                                                            1\pm \alpha(\nu,k) & \beta(\nu,k) \\
                                                                             \beta(\nu,k)& 1\mp \alpha(\nu,k) \\
                                                                           \end{array}
                                                                         \right)\,,\label{ppmlast} \ee and

\bea
 M^2(\nu,k) & = &  -(\nu-i\epsilon)^2 +  \frac{1}{2}\Big[\omega^2_1(k)+\omega^2_2(k)   +(g^2_1+g^2_2)\,\sigma(\nu,k) \Big]\,, \label{M2last}\\
\mathcal{D}(\nu,k) & = &  \Bigg[\Big(m^2_1 -m^2_2 + (g^2_1-g^2_2)\,\sigma(\nu,k)\Big)^2+4 g^2_1\,g^2_2\,\sigma^2(\nu,k)   \Bigg]^{1/2}\,, \label{Dlast}\\
\alpha(\nu,k) & = &  \frac{1}{\mathcal{D}(\nu,k)}\,\Big[ m^2_1 -m^2_2 +  (g^2_1-g^2_2)\,\sigma(\nu,k)  \Big]\,,\label{alfilast}\\   \beta(\nu,k) & =  & \frac{2 \,g_1\, g_2\,\sigma(\nu,k)}{\mathcal{D}(\nu,k)}  \,. \label{betilast}
\eea
In the above expressions, $\omega_{1,2}(k)$ are the renormalized frequencies and  $\sigma(\nu,k)$ is the \emph{renormalized} self-energy (without the couplings), it is given by eqn. (\ref{sigreim})  after subtracting the zero temperature ultraviolet divergent part\footnote{An alternative renormalization scheme would keep the ultraviolet finite zero temperature parts.}.
 The final result is obtained by complex integration by closing the contour in the upper half complex $\nu$-plane (for $t>0$), and recognizing the position of complex poles and/or branch cut singularities, which necessarily depend on the values of masses and couplings.

 \section{Dynamics of neutral pion-axion mixing}\label{sec:piondynamics}
 The formulation and results above are general for scalar and/or pseudoscalar fields interacting indirectly through their coupling to a common bath in equilibrium and applies to any case of pseudoscalar meson mixing with axions or among different axion-like species. In this section we focus on the particular example of neutral pion-axion mixing after the QCD phase transition, under the following main assumptions: \textbf{i:)} the neutral pion mass  is much larger than the axion mass $m_\pi \gg m_a$, \textbf{ii:)} the neutral pion coupling to photons is much larger than that of the axion, namely $g_1\gg g_2$, this latter assumption is motivated by the axion being assumed to be a very weakly interacting dark matter candidate with a lifetime of the order of the Hubble time or larger. Under these assumptions it follows that
 \bea \mathcal{D}(\nu,k) & \simeq &  \Big(m^2_1 -m^2_2 + (g^2_1-g^2_2)\,\sigma(\nu,k)\Big)\,\Bigg[1 + \frac{2 g^2_1 g^2_2 \sigma^2(\nu,k)}{(m^2_1-m^2_2)^2}+\cdots \Bigg]\,, \label{Dapp}\\\alpha(\nu,k) & \simeq & 1- \frac{2 g^2_1 g^2_2 \sigma^2(\nu,k)}{(m^2_1-m^2_2)^2}  \,,\label{alfaap}\\ \beta(\nu,k) & \simeq & \frac{2 g_1 g_2 \sigma(\nu,k)}{(m^2_1-m^2_2)} \equiv \varepsilon(\nu,k)  \,. \label{betap}
 \eea  Keeping only terms up to order $g^2_1,g^2_2, g_1g_2$, consistently with obtaining the effective action up to second order in couplings,  we find
 \bea &&  M^2(\nu,k)+ \frac{\mathcal{D}(\nu,k)}{2} \simeq -(\nu-i\epsilon)^2+\omega^2_1(k)+g^2_1\,\sigma(\nu,k)\,,  \label{mpd}\\ &&  M^2(\nu,k)- \frac{\mathcal{D}(\nu,k)}{2} \simeq -(\nu-i\epsilon)^2+\omega^2_2(k)+g^2_2\,\sigma(\nu,k)\,,  \label{mmd}\\&& \mathbb{P}_+(\nu,k) \simeq \left(
                                              \begin{array}{cc}
                                                1 & \frac{\varepsilon(\nu,k)}{2} \\
                                                \frac{\varepsilon(\nu,k)}{2} &0
                                              \end{array}
                                            \right)\,,\label{ppap}\\
&& \mathbb{P}_-(\nu,k) \simeq \left(
                                              \begin{array}{cc}
                                                0 & -\frac{\varepsilon(\nu,k)}{2} \\
                                                -\frac{\varepsilon(\nu,k)}{2} &1
                                              \end{array}
                                            \right)\,.\label{pmap} \eea

                       In this limit, the two contributions to the Green's function (\ref{goft2}) have a simple interpretation: the term with $\mathbb{P}_+$ is identified with a pion-like pole and that with $\mathbb{P}_-$ with the
                       axion-like pole.

In the Breit-Wigner approximation, these contributions feature complex poles in the upper half $\nu$-plane at
\be \nu_{1,2}(k) = \pm \omega_{1,2} + i \, \frac{\Gamma_{1,2}(k)}{2} \,\label{polos} \ee   for pion-like (1) or axion-like (2) respectively, with
\be  \Gamma_{1,2}(k) = g^2_{1,2}\,\frac{\sigma_I(\omega_{1,2}(k))}{\omega_{1,2}(k)}\,.\label{widths12}\ee The integral in (\ref{goft2}) can be carried out in this approximation, yielding the Green's function (in matrix form)
\bea \mathcal{G}(k;t) & = & Z_1\,e^{-\frac{\Gamma_1(k)}{2}t}\,\frac{\sin(\omega_1(k)t)}{\omega_1(k)}\,\left(
                                                                                            \begin{array}{cc}
                                                                                              1 & 0 \\
                                                                                             0 & 0 \\
                                                                                            \end{array}
                                                                                          \right)+
 Z_2\, e^{-\frac{\Gamma_2(k)}{2}t}\,\frac{\sin(\omega_2(k)t)}{\omega_2(k)}\,\left(
                                                                                            \begin{array}{cc}
                                                                                              0 & 0 \\
                                                                                             0 & 1 \\
                                                                                            \end{array}
                                                                                      \right)\nonumber \\ & + &
\Big(Z_1 \,\mathcal{F}_1(t)- Z_2\,\mathcal{F}_2(t)\Big)\,\left(
                                                                                            \begin{array}{cc}
                                                                                              0 & 1 \\
                                                                                             1 & 0 \\
                                                                                            \end{array}
                                                                                      \right)\,,\label{Goftfins}\eea                                                                                          where for  $a=1,2$,
\be Z_a = \Big[1- g^2_a\,\frac{\partial\,\sigma_R(\nu,k) }{\partial \nu}\Bigg|_{\nu=\omega_a(k)}\Big]^{-1}\,,\label{Zetas} \ee are the wave function renormalization constants which are finite after cancelling the ultraviolet divergences of the self-energy with proper counterterms, and
\be \mathcal{F}_{a}(t) = e^{-\frac{\Gamma_a(k)}{2}t}\, \Big(\gamma_{a+}\,   \frac{e^{i\omega_a(k) t}}{2i\omega_a(k)}- \gamma_{a-}  \,\frac{e^{-i\omega_a(k) t}}{2i\omega_a(k)}\Big) ~~;~~ \gamma_{a\pm} = \frac{g_1g_2}{m^2_1-m^2_2}\,\Big(\sigma_R(\omega_a(k))\pm i \sigma_I(\omega_a(k) \Big)\,.\label{bigFs}\ee

In the following analysis we will assume that the wavefunction renormalization constants $Z_a$ are absorbed into the usual field redefinitions by adding proper diagonal counterterms in the Lagrangian so that the diagonal part of    poles in the Green's functions feature residue equal to one, thereby setting $Z_1=Z_2=1$ in (\ref{Goftfins}).

Armed with this result, we can now study the evolution of expectation values, correlation functions and populations, by implementing the general results (\ref{conncorrfin},\ref{eneA},\ref{Asplits},\ref{realtisol}).

\vspace{2mm}

\subsection{Induced pion condensate}\label{subsec:picond}
The expectation value $\overline{\langle \langle \Phi(\vk,t) \rangle \rangle}$  is determined solely by the homogeneous solution in eqn. (\ref{Asplits}) and given by   eqn. (\ref{homog}), because the average over the noise vanishes (see eqn. (\ref{noiscors})). Let us consider that initially there is a misaligned axion condensate with vanishing velocity, and no pion condensate, namely
\be \Phi_{i}(\vk) = \left(
                      \begin{array}{c}
                        0 \\
                        a_i(\vk) \\
                      \end{array}
                    \right)~~;~~ \Pi_{i}(\vk) = \left(
                      \begin{array}{c}
                        0 \\
                        0  \\
                      \end{array}
                    \right)\,, \label{initialcond}\ee  yielding

\be\overline{\langle \langle \Phi(\vk,t) \rangle \rangle} = \left(
                      \begin{array}{c}
                        \overline{\pi^0}(\vk,t) \\
                        \overline{a}(\vk,t) \\
                      \end{array}
                    \right)~  \ee
with
\be   \overline{a}(\vk,t) = a_i(\vk) \,e^{-\frac{\Gamma_2(k)}{2}t}\,\cos(\omega_2(k)t)\,, \label{expecat}\ee   where we neglected terms of order $\Gamma_a/\omega_a(k) \ll 1$.
The off-diagonal components of the Green's function (\ref{Goftfins})   induce an expectation value of the pion field, namely a  pion condensate $\overline{\pi^0}(\vk,t)$ , which is given by
 \be \overline{\pi^0}(\vk,t)= \frac{a_i(\vk)}{2}\,\Big[\gamma_{1+}\,e^{-\frac{\Gamma_1(k)}{2}t}\,e^{i\omega_1(k)t}\,+\,\gamma_{1-}\,e^{-\frac{\Gamma_1(k)}{2}t}\,e^{-i\omega_1(k)t} -\,\gamma_{2+}\,e^{-\frac{\Gamma_2(k)}{2}t}\,e^{i\omega_2(k)t}-\,\gamma_{2-}\,e^{-\frac{\Gamma_2(k)}{2}t}\,e^{-i\omega_2(k)t} \Big]\,,\label{picond}  \ee this result confirms the conjecture in ref.\cite{shuyangcs}: a misaligned axion condensate induces a pion condensate. The expression (\ref{picond}) is noteworthy, it is a linear combination of a short-lived (pion-like) component decaying on a time scale $1/\Gamma_1$ and a long-lived (axion-like) component decaying on a time scale $1/\Gamma_2 \gg 1/\Gamma_1$. In this sense, the result (\ref{picond})    is reminiscent of the time  evolution of neutral kaon beams where the propagation eigenstates correspond to   short lived and   long-lived ($K_s,K_L$) states. An initially prepared $K^0$  beam (from a strong interaction reaction)  evolves in time as a linear superposition of the short and the long lived states and   at times longer than the short lived lifetime but shorter than the long-lived component it  ``purifies'' into the long lived component. A similar dynamics is explicit in the solution (\ref{picond}), despite the fact that neutral pions feature a lifetime $1/\Gamma_1(k)$ much shorter than that of axions, namely $1/\Gamma_2(k)$, their indirect coupling  induces a long-lived pion condensate. During the window of time $1/\Gamma_1(k) \ll t \ll 1/\Gamma_2(k)$ and to leading order in the couplings, the induced pion condensate becomes ``purified'' in the sense that it is  solely determined by the long-lived  misaligned axion condensate
 \be \overline{\pi^0}(\vk,t)= -\frac{g_1}{g_2(m^2_1-m^2_2)}\,\Bigg[\Sigma^R_{ax}(k)\,\overline{a}(\vk,t) + \Gamma_{ax}(k) \,\dot{\overline{a}}(\vk,t) \Bigg]\,,\label{puripi}\ee with the axion   self-energy and decay rate
 \be \Sigma^R_{ax}(k)= g^2_2 \,\sigma_R(\omega_2(k),k)~~;~~\Gamma_{ax}(k)=g^2_2 \, \frac{\sigma_I(\omega_2(k),k)}{\omega_2(k)}\,.\label{axs}  \ee We note that the expression in the bracket in (\ref{puripi}) is the induced abelian $U(1)$ Chern-Simons density found in ref.\cite{shuyangcs}, namely
 \be \langle \vec{E}(t)\cdot \vec{B}(t) \rangle = -\frac{1}{g_2}\,\Bigg[\Sigma^R_{ax}(k)\,\overline{a}(\vk,t) + \Gamma_{ax}(k) \,\dot{\overline{a}}(\vk,t) \Bigg]\,.\label{cs}\ee
  Therefore, eqn. (\ref{puripi})  confirms one of the main conjectures   in that reference, namely that a misaligned axion condensate induces a neutral pion condensate proportional to the Chern Simons condensate.

  It is important to highlight differences with the Kaon system: in absence of coupling to the common set of intermediate states (weak interactions in the Kaon case), $K^0,\overline{K}^0$ are degenerate. Therefore their mass splitting as a consequence of mixing is of the same order as their decay widths, hence, Kaon dynamics features \emph{interference} terms manifest as oscillations. Obviously these are not relevant in the case of axion-pion mixing because of the large difference in masses that suppresses the interference terms, averaging them out on very short time scales. Nevertheless, this important difference with the axion-neutral pion case notwithstanding, the physical reason for the ``purification'' is the same, a large discrepancy in the decay lifetimes of the propagating states.

 It is clear from equation (\ref{picond}) that the amplitude of the induced pion condensate   is very small as it is proportional to the axion-photon coupling. However, near the QCD phase transition associated with chiral symmetry breaking and the binding of quarks into  the pseudoscalar bound states there is an instability associated with this symmetry breaking, such instability will amplify the ``seed'' induced by the axion field given by (\ref{picond}). Therefore, even for a small induced condensate ``seed'' the instability towards chiral symmetry breaking results in an amplification of its initial value. It is not the purpose of this study to assess the phenomenology of chiral symmetry breaking but to highlight the fundamental result that the ``seeding'' of the pion condensate by a misaligned axion is a direct and  unambiguous consequence of axion-pion mixing revealed by  the non-equilibrium framework introduced above. The full non-equilibrium dynamics of the pion condensate during the transition to the lower free energy state including the concomitant instabilities must be addressed within a more complete theory of   chiral symmetry breaking, eqn. (\ref{picond}) describes only the ``seeding'' of the neutral pion condensate.

\subsection{Connected correlation functions:}\label{subsec:corres}
The connected correlation functions are defined by eqn. (\ref{conncorrfin}), there are two distinct contributions, one from the homogeneous solutions $\Phi^h$ (\ref{homog})  and another from the noise dependent part of the solution $\Phi^{\xi}$ (\ref{realtisol}), writing the total correlation function  (\ref{conncorrfin}) in obvious notation as
\be \langle \phi_a(\vk,t)\phi_b(-\vk,t)\rangle_c \equiv \mathcal{I}^{h}_{ab}(t)+\mathcal{I}^{\xi}_{ab}(t)\,, \label{corresplit}\ee the homogeneous contribution is given by
\be \mathcal{I}^{h}_{ab}(k;t) = \dot{\mathcal{G}}_{ac}(t)\,\overline{\Phi_{c,i}(\vk)\Phi_{d,i}(\vk)}\, \dot{\mathcal{G}}_{db}(t)+ {\mathcal{G}}_{ac}(t)\,\overline{\Pi_{c,i}(\vk)\Pi_{d,i}(\vk)}\,  {\mathcal{G}}_{db}(t)+\mathrm{crossed~terms}\,. \label{homocorr}\ee because each Greens functions features a decaying exponential the homogeneous contribution decays exponentially, thereby erasing memory of the initial conditions.

We now consider the noise contribution, for which we need   the noise correlation function (\ref{noiscors}) with the  spectral representation (\ref{noisedis}) shown explicitly in appendix (\ref{app:correlations}), namely
\bea \mathcal{N}_{ab}(k,t-t') & = &  \mathbb{C}_{ab}\,\int N(k_0,k)\,e^{-ik_0(t-t')}\,\frac{dk_0}{2\pi}\, \label{nabo}\\
N(k_0,k) & = & \frac{\rho(k_0,k)}{2}\,\coth\Big[\frac{\beta k_0}{2}\Big]\,,\label{nkcero}\eea along with the inhomogeneous solution (\ref{realtisol}). Using that  the Greens function matrix is symmetric, the noise contribution to the connected correlation function (\ref{conncorrfin}) becomes
\be \mathcal{I}^{\xi}_{ab}(t)   = \int^{\infty}_{-\infty} N(k_0,k) \Big[\int^t_0 \mathcal{G}_{ac}(\tau)\,e^{ik_0\tau} \Big]\,\mathbb{C}_{cd}\,\Big[\int^t_0 \mathcal{G}_{db}(\tau)\,e^{-ik_0\tau} \Big]\,\frac{dk_0}{2\pi}\,.\label{corrcorr} \ee Each integral is straightforward to obtain, their product feature direct terms of the form
\be \frac{1}{k_0\pm \omega_a + i\,\frac{\Gamma_a}{2}}\,\frac{1}{k_0\pm \omega_a - i\,\frac{\Gamma_a}{2}}\,,\label{direct}\ee indirect terms of the form
\be \frac{1}{k_0\pm \omega_a + i\,\frac{\Gamma_a}{2}}\,\frac{1}{k_0\mp  \omega_a - i\,\frac{\Gamma_a}{2}}\,,\label{indirect}\ee and interference terms that mix different components. The integral over $k_0$ is dominated by the various poles in the complex $k_0$-plane, the ones just described, and those at the Matsubara frequencies $k_0 = 2\pi n/\beta~;~ n=  \pm 1,\pm 2\cdots$, the $n=0$ term features vanishing residue because $\rho(0,k)=0$. The direct terms feature resonant poles with residues $\propto 1/\Gamma_a \propto 1/g^2_a$, the indirect terms are non-resonant featuring residues of the form $1/(\omega_a \pm i\Gamma_a/2)$, therefore subleading with respect to the direct terms since $\omega_a \gg \Gamma_a$, with a similar behavior for the interference terms featuring denominators of the form $1/(\omega_1-\omega_2 +i (\Gamma_1-\Gamma_2))$ with $\omega_1 \gg \omega_2$, and  the poles at the Matsubara frequencies, therefore the $k_0$ integral is dominated by the resonant poles. A long but straightforward calculation finally yields
\bea \mathcal{I}^{\xi}(k;t)  & = & \frac{1+2\,n(\omega_1)}{2\omega_1}\,(1-e^{-\Gamma_1 t})\,
\Big(1+\mathcal{O}(g^2_{1,2})\Big) \left(
                                \begin{array}{cc}
                                  1 & 0 \\
                                  0 & 0 \\
                                \end{array}
                              \right) \nonumber \\
& + &  \frac{1+2\,n(\omega_2)}{2\omega_2}\,(1-e^{-\Gamma_2 t})\,
\Big(1+\mathcal{O}(g^2_{1,2}) \Big) \left(
                                \begin{array}{cc}
                                  0 & 0 \\
                                  0 & 1 \\
                                \end{array}
                              \right) \nonumber \\
& + &      g_1 g_2 \Bigg[\frac{1+2\,n(\omega_1)}{2\omega_1}\,(1-e^{-\Gamma_1 t})\,\Delta_1\,-\, \frac{1+2\,n(\omega_2)}{2\omega_2}\,(1-e^{-\Gamma_2 t})\,\Delta_2 \Bigg]\left(
                                                                         \begin{array}{cc}
                                                                           0 & 1 \\
                                                                           1 & 0 \\
                                                                         \end{array}
                                                                       \right)  \label{concorfi}\eea
 where
  \be \Delta_a= \frac{\sigma_R(\omega_a(k))}{m^2_1-m^2_2}\,,\label{deltaas} \ee
   used the relations (\ref{widths12}) and kept the leading order terms in the couplings. Whereas the homogeneous contribution (\ref{homocorr}) depends on the initial conditions, the noise contribution (\ref{concorfi}) does not and is solely a consequence of the bath.

     In order to understand the combined results (\ref{homocorr},\ref{concorfi}), let us compare to the equal time correlation function of the free scalar fields in thermal equilibrium,
\be \langle \phi_a(\vk,t)\phi_b(-\vk,t)\rangle = \frac{1+2\,n(\omega_1)}{2\omega_1} \left(
                                                                                      \begin{array}{cc}
                                                                                        1 & 0 \\
                                                                                        0 & 0 \\
                                                                                      \end{array}
                                                                                    \right)
+ \frac{1+2\,n(\omega_2)}{2\omega_2} \left(
                                                                                      \begin{array}{cc}
                                                                                        0& 0 \\
                                                                                        0 & 1 \\
                                                                                      \end{array}
                                                                                    \right)
\,.\label{equicor}\ee

On long time scales $t \gg 1/\Gamma_1, 1/\Gamma_2$ the mixing fields approach thermal equilibration with the common bath. The off-diagonal elements are a distinct signature of the long-lived correlations as a consequence of mixing. We note that the off diagonal correlations that survive in the long time limit $t \gg 1/\Gamma_{1,2}$ are independent of the initial conditions and arise solely from the noise contribution, namely the bath, because the homogeneous contribution (\ref{homocorr}) vanishes in this limit.

Repeating the calculation for the populations (\ref{eneA}), keeping solely the connected contribution and neglecting quadratic corrections  in the couplings we find
\be n_a(k;t) = n(\omega_a)\,(1+\mathcal{O}(g^2_{1,2}))\,(1-e^{-\Gamma_a(k)t}) \,,\label{popus} \ee which clearly indicates the approach to thermalization with the common bath. For the case of axions only, this result confirms those of ref.\cite{shuyang}.

The  results (\ref{homocorr},\ref{equicor}) are noteworthy, the off-diagonal correlations, namely the coherences as defined above, are a distinct consequence of the indirect coupling mediated by the bath. This is a manifestation of bath (or environment) induced coherence between the mixing fields. While the homogeneous contribution (\ref{homocorr}) vanishes in the long time limit erasing the memory of the initial conditions, the off-diagonal contribution from the noise remains non-vanishing in this limit.  Note that this contribution to the coherence is established on the shortest time scale, namely, even if the axion field thermalizes on a very long time scale,  the thermalization of the pion field leads to non-vanishing coherence that is long-lived and remains even beyond the axion lifetime.

\section{Summary of results, conclusions,   and further questions}\label{sec:conclusions}

In this study we introduced a framework to study the non-equilibrium dynamics of \emph{mixing} of scalar or pseudoscalar fields that each couples to a common bath of degrees of freedom in thermal equilibrium.  Field mixing is a result of this   coupling to a set of common intermediate states that results in off-diagonal self-energies.  Our motivation for this study stems from the observation in ref.\cite{shuyangcs} that   axions can mix with a neutral pion after the QCD phase transition via a common two photon channel, since they both couple to this channel via the Chern-Simons density, a consequence of the U(1) triangle anomaly. Furthermore, it was conjectured in this reference that a misaligned axion can \emph{induce} a pion condensate and that such macroscopic condensate is proportional to a condensate of the Chern-Simons density, itself induced by the misaligned axion.

The non-equilibrium dynamics of field mixing in a medium is,  however, of much broader fundamental interest, encompassing the possibility of other axion-like particles in extensions beyond the standard model mixing with    axions, or in extensions that feature ``portals'' connecting the standard model to extensions beyond it, as such portals may provide intermediate states that could indirectly mix both sides through their mutual coupling to these ``portal states''. A framework that extends the formulation of meson mixing in \emph{vacuum} to the case when the intermediate states are in a medium, may prove relevant   to study CP violation and/or baryogenesis in cosmology, since flavored meson mixing in vacuum yields direct and distinct observational signatures of CP violation. Furthermore, this study was also motivated by the tantalizing possibility that axion-like degrees of freedom are emergent quasiparticles in a wide range of condensed matter systems, from topological insulators and Weyl semimetals to multilayered metamaterials or multiferroics also coupling to electromagnetism via the Chern-Simons density.

In this article we obtained  the non-equilibrium effective action that determines the time evolution of the reduced density matrix for the pseudoscalar fields by extending and generalizing the methods introduced in ref.\cite{shuyang,boyqbm} to the case of field mixing, and applied this framework to study axion-neutral pion mixing as a specific example from which we can draw broader lessons.

\textbf{Summary of results:} Although the neutral pion and the axion feature widely different masses and decay widths, the non-equilibrium mixing dynamics displays a wealth of interesting phenomena.

\begin{itemize}
\item{ A misaligned axion condensate \emph{induces} a macroscopic condensate of the neutral pion thereby confirming a conjecture in ref.\cite{shuyangcs}. The pion condensate exhibits evolution on a long and a short time scale akin to the $K^0-\overline{K}^0$ in vacuum, this induced pion condensate survives on time scales much longer than the pion lifetime and eventually decays on the much longer axion-like time scale. This phenomenon is reminiscent of the ``purification'' into a long-lived Kaon state in the $K^0-\overline{K}^0$ system, albeit with important differences because the axion-neutral pion system features a large mass difference  in the absence of mixing, thereby suppressing interference effects.}

    \item{On time scales much longer than the pion lifetime, the neutral pion condensate is proportional to the macroscopic condensate of the Chern-Simons density found in ref.\cite{shuyangcs}, confirming another conjecture in this reference. Taken together, these results indicate that a misaligned axion condensate does induce a macroscopic neutral pion condensate after the QCD phase transition. If the lightest up and down quarks were massless, this condensate would imply chiral symmetry breaking, hence the axion  ``seeds'' chiral symmetry breaking in QCD as a consequence of its indirect mixing with the neutral pion via the U(1) anomaly.
        While the amplitude of the induced pion condensate is small, being proportional to the axion-photon coupling, the instability associated with the chiral phase transition in QCD will amplify this small seed.  }

        \item{Another important consequence of mixing, is that the effective field theory obtained upon tracing out the bath degrees of freedom \emph{must} include kinetic mixing terms with higher derivative operators. This is a consequence of the non-renormalizability of the effective field theory, and manifest in ultraviolet divergences in the zero temperature contribution of the one loop self-energies which are quadratic and \emph{quartic} in the four momenta.}

  \item{The off-diagonal components of the axion-pion connected correlation functions, interpreted as \emph{coherences} in analogy with the density matrix of two level systems (qubits), exhibit thermalization and remain non-vanishing and independent of initial conditions  even on the longer time scales of axion decay. This is a manifestation of long-lived bath induced coherences, a consequence of  the indirect interaction of pions and axions mediated by the common bath. }

\end{itemize}

While the effects of axion-neutral pion mixing are undoubtedly rather small, because the axion is very weakly coupled as befits a dark matter candidate, the main framework and many results are of fundamental relevance, a ``proof of principle''  and hitherto unexplored. The effective action for mixing in a medium has a far broader applicability than the specific example studied here. As mentioned above, it may prove important in complementary studies of the non-equilibrium dynamics of CP violation and/or baryogenesis in cosmology, after all flavored meson mixing in vacuum is one of the observational pillars of CP violation, and perhaps in ``portal'' extensions beyond the standard model. Several of the results obtained for axion-neutral pion mixing transcend this example:  if one of the mixing fields features a non-vanishing expectation value (condensate), a condensate of the other   field will be induced as a consequence of mixing. An off-diagonal self energy, a hallmark of indirect mixing, will also result in non-vanishing off diagonal correlations, namely ``bath-induced'' coherence  between the fields, even when initially they are uncorrelated. An important aspect that merits further study is the dynamics under conditions of near degeneracy, namely when the mass difference is of the same order as the differences in the widths. Under this circumstance interference effects will become important and must be included in the dynamics. We expect to report on such study in future work.

\acknowledgements
  The authors gratefully acknowledge  support from the U.S. National Science Foundation through grant   NSF 2111743.

\appendix

\section{Lehmann representation of correlation functions}\label{app:correlations}
The correlation functions $G^>(x-y);G^<(x-y)$ can be written in an exact Lehmann (spectral) representation which is useful to include in the equations of motion.
\bea G^> (x-y) & = & \frac{1}{Z_\chi}\mathrm{Tr} e^{-\beta H_{\chi}} \mathcal{O} (x)\mathcal{O} (y)\, \label{ggreatap} \\ G^< (x-y) & = & \frac{1}{Z_\chi}\mathrm{Tr} e^{-\beta H_{\chi}} \mathcal{O} (y)\mathcal{O} (x)\,, \label{glessap}\eea
where $Z_{\chi} = \mathrm{Tr} e^{-\beta H_{\chi}}$, $\mathcal{O}(\vx,t) = \vec{E}(\vx,t)\cdot \vec{B}(\vx,t)$ and
$\mathcal{O} (\vx,t)= e^{iH_{\chi}t}\,e^{-i\vec{P}\cdot \vx}\, \mathcal{O} (0)\,e^{i\vec{P}\cdot \vx}\,e^{-iH_{\chi}t}$. In terms of a complete set of simultaneous eigenstates of $H_{\chi},\vec{P}$, namely $(H_{\chi},\vec{P})|n\rangle = (E_n,\vec{P}_n)|n\rangle $ and inserting the identity in this basis, we find
\bea   G^> (x_1-x_2)  &  = &   \frac{1}{Z_{\chi}}{\sum_{n,m}} e^{-\beta E_n} e^{i(E_n-E_m)(t_1-t_2)}\,e^{-i(\vec{P}_n-\vec{P}_m)\cdot(\vx_1-\vx_2)}\,\langle n|\mathcal{O} (0)|m\rangle \langle m|\mathcal{O} (0)|n\rangle \,, \nonumber \\\label{ggreatrep}\\
G^< (x_1-x_2)  &  = &   \frac{1}{Z_{\chi}}{\sum_{n,m}} e^{-\beta E_n} e^{-i(E_n-E_m)(t_1-t_2)}\,e^{i(\vec{P}_n-\vec{P}_m)\cdot(\vx_1-\vx_2)}\,\langle n|\mathcal{O}(0)|m\rangle \langle m|\mathcal{O}(0)|n\rangle \,. \nonumber \\ \label{glessrep}
\eea
 These representations may be written in terms of spectral densities, by introducing
 \bea && \rho^> (k_0,\vk)  =   \frac{(2\pi)^4}{Z_{\chi}}{\sum_{n,m}} e^{-\beta E_n}  \,\langle n|\mathcal{O} (0)|m\rangle \langle m|\mathcal{O} (0)|n\rangle \, \delta(k_0 - (E_m-E_n)) \delta^{3}(\vk- (\vec{P}_m-\vec{P}_n))  \,, \nonumber \\ \label{rhogreatrep}\\
&& \rho^<(k_0,\vk)     =     \frac{(2\pi)^4}{Z_{\chi}}{\sum_{n,m}} e^{-\beta E_n}  \,\langle n|\mathcal{O}(0)|m\rangle \langle m|\mathcal{O}(0)|n\rangle  \delta(k_0 - (E_n-E_m)) \delta^{3}(\vk- (\vec{P}_n-\vec{P}_m))\,,\nonumber  \\ \label{rholessrep}
\eea in terms of which
\bea G^>(x_1-x_2)  &  = & \int \frac{d^4k}{(2\pi)^4}\,  \rho^>(k_0,\vk)\, e^{-ik_0(t_1-t_2)}\,e^{i\vk\cdot(\vx_1-\vx_2)} \,\label{specgreat} \\
 G^<(x_1-x_2)  &  = & \int \frac{d^4k}{(2\pi)^4}\,  \rho^<(k_0,\vk)\, e^{-ik_0(t_1-t_2)}\,e^{i\vk\cdot(\vx_1-\vx_2)} \,.\label{specless} \eea   Relabelling $n \leftrightarrow m$ and using the $k_0$ delta function   in (\ref{rholessrep}), we find the   Kubo-Martin-Schwinger condition\cite{kms}
 \be  \rho^<(k_0,\vk) = e^{-\beta k_0}\, \rho^>(k_0,k)\,. \label{kmscond} \ee Introducing the spectral density
 \be  \rho (k_0,\vk)=  \rho^> (k_0,\vk)- \rho^< (k_0,\vk)= \rho^> (k_0,\vk)\,\big(1-e^{-\beta k_0}\big) \,,\label{spectral}\ee it follows that
 \be \rho^> (k_0,\vk)= \big(1+ n(k_0) \big) \rho (k_0,\vk)~~;~~ \rho^< (k_0,\vk)=  n(k_0) \, \rho (k_0,\vk)\,,\label{relas}\ee
 where
 \be n(k_0) = \frac{1}{e^{\beta k_0}-1}\,. \label{nofk0}\ee
 Therefore the spatial Fourier transform of the self-energy matrix (\ref{kernelsigma}) and the noise kernel (\ref{kernelkappa})  can be written as
 \bea \Sigma_{ab}(k;t-t')& = & -i \, \mathbb{C}_{ab}\int \frac{dk_0}{(2\pi)}\,\rho (k_0,k)  e^{-ik_0(t-t')}  \label{sigmadis} \\ \mathcal{N}_{ab}(k;t-t') & = & \frac{1}{2}\,\mathbb{C}_{ab}\,\int \frac{dk_0}{(2\pi)}\,\rho (k_0,k)\, \coth\big[ \frac{\beta k_0}{2}\big]\,  e^{-ik_0(t-t')} \,, \label{noisedis} \eea this is the general relation between the self-energy and the noise correlation function commonly determined by the spectral density $\rho(k_0,k)$, a direct consequence of the fluctuation-dissipation relation as a result of the bath being in thermal equilibrium, namely in terms of the time-Fourier transform and in obvious notation,
 \be  2\,\mathcal{N}_{ab}(k_0;k) = i \Sigma_{ab}(k_0;k)\,\coth\big[ \frac{\beta k_0}{2}\big]\,.\label{fdr}  \ee

 For the case under consideration $\mathcal{O}(\vx,t) = \vec{E}(\vx,t)\cdot \vec{B}(\vx,t)$ and with the thermal ensemble (radiation bath in equilibrium with charged fields) being invariant under rotations, it follows that $\rho(k_0,\vk) = \rho(k_0,k)$ and the relations above lead to the property
 \be \rho(k_0,k) = - \rho(-k_0,k)\,. \label{rhoodd} \ee

 These results are valid to all orders in the couplings of the electromagnetic field to any charged field within or beyond the standard model, including charged particle  loop corrections to the internal photon propagators in the self-energy.

 Obviously the full spectral density is not available, however the one free-photon loop contribution has been obtained in ref.\cite{shuyang}, it is given by
 \be \rho(k_0,k) = \rho^{(0)}(k_0,k)+\rho^{(T)}(k_0,k)\,,\label{rho0T}\ee where the zero $(0)$ and finite temperature $(T)$ parts are given by
 \bea \rho^{(0)}(k_0,k) & = &   \frac{(K^2)^2}{32\pi}\,\Theta(K^2)\,\mathrm{sign}(k_0)\,,\label{rhozero} \\
 \rho^{(T)}(k_0,k) & = &  \frac{(K^2)^2 }{16\pi\,\beta\,k}\,\Bigg\{   \ln\Bigg[\frac{1-e^{-\beta \omega^I_+}}{1-e^{-\beta \omega^I_-}} \Bigg] \,\Theta(K^2)  +   \ln\Bigg[\frac{1-e^{-\beta \omega^{II}_+}}{1-e^{-\beta \omega^{II}_-}} \Bigg]\,\Theta(-K^2) \Bigg\}\, \mathrm{sign}(k_0)\,,\nonumber \\
&& \beta = \frac{1}{T}~~;~~ K^2=k^2_0-k^2 ~~;~~   \omega_\pm^{(I)}   =   \frac{|k_0| \pm k}{2}~~;~~
    {\omega}_\pm^{(II)} = \frac{k \pm |k_0|}{2}\,,\label{rhoT} \eea  which explicitly shows the property (\ref{rhoodd}).
  The terms with $\Theta(k^2_0-k^2)$ arise from the processes $\phi_{1,2} \leftrightarrow 2 \gamma$, namely emission and absorption of photons with the reverse or recombination process $2\gamma \rightarrow \phi_{1,2}$  being a consequence of the radiation bath, these processes  feature support on the axion and pion mass shells for massive axions. The contribution proportional to $\Theta(k^2-k^2_0)$ only features support below the light cone, it describes off-shell processes $\gamma \phi_{1,2} \leftrightarrow \gamma$ and vanishes in the $k\rightarrow 0$ limit.

\section{Green's function}\label{app:green}

Let us write
\bea
\M_{11} & = &  s^2+k^2+m^2_1+\Delta m^2_{11}+\widetilde{\Sigma}_{11}(\vk,s)\nonumber \\
\M_{12} & = &   \Delta m^2_{12}+\widetilde{\Sigma}_{12}(\vk,s)~~;~~ \M_{21}   =     \Delta m^2_{21}+\widetilde{\Sigma}_{21}(\vk,s)\nonumber \\
\M_{22} & = &  s^2+k^2+m^2_2+\Delta m^2_{22}+\widetilde{\Sigma}_{22}(\vk,s)\,,\label{elemts}
\eea
in terms of which the general form of $\mathbb{G}^{-1}_{ab}(k,s)$ in eqn. (\ref{inverG}) is written as
\be \mathbb{G}^{-1}_{ab}(k,s) =  \left(
                                   \begin{array}{cc}
                                     \M_{11} & \M_{12} \\
                                     \M_{21} & \M_{22} \\
                                   \end{array}
                                 \right) \equiv M^2\,\mathbf{1}+\frac{\D}{2}\,\left(
                                                                           \begin{array}{cc}
                                                                             \alpha & \beta \\
                                                                             \gamma & -\alpha \\
                                                                           \end{array}
                                                                         \right) \,,\label{invG}
\ee
where, neglecting the labels $k,s$ to simplify notation, which will be assumed as arguments in the quantities defined below, we introduced
\bea
\D & = & \Bigg[\Big(\M_{11}-\M_{22}\Big)^2+ 4\,\M_{12}\,\M_{21} \Bigg]^{1/2} \label{det}\\
M^2 & = & \frac{1}{2}\Big(\M_{11}+\M_{22} \Big)~~;~~ \alpha= \frac{\Big(\M_{11}-\M_{22} \Big)}{\D} \nonumber \\
\beta & = & \frac{2\,\M_{12}}{\D}~~;~~ \gamma = \frac{2\,\M_{21}}{\D} \label{mtxelet}
\eea from which it follows that
\be \alpha^2+\beta \gamma =1 \,, \label{idal}\ee therefore
\be \mathrm{det} \left(
                                                                           \begin{array}{cc}
                                                                             \alpha & \beta \\
                                                                             \gamma & -\alpha \\
                                                                           \end{array}
                                                                         \right) = -1 \,\label{deter}\ee hence, the eigenvalues of this matrix are $\pm 1$, and
 \be \mathrm{det}[\mathbb{G}^{-1}] = \Big(M^2 + \frac{\D}{2}\Big)\,\Big(M^2 - \frac{\D}{2} \Big) \,.     \label{detinvG}\ee      The inverse of the matrix (\ref{invG}) is given by
\be \mathbb{G} = \frac{1}{\mathrm{det}[\mathbb{G}^{-1}] }\, \Bigg[ M^2 \mathbf{1}- \frac{\D}{2}\,\left(
                                                                           \begin{array}{cc}
                                                                             \alpha & \beta \\
                                                                             \gamma & -\alpha \\
                                                                           \end{array}
                                                                         \right) \Bigg]\,. \label{Gifi}\ee
 Writing
 \bea
 M^2 & = &  \frac{1}{2}\,\Big( M^2 + \frac{\D}{2} \Big)+ \frac{1}{2}\,\Big( M^2 - \frac{\D}{2} \Big)\,, \nonumber \\
 \D & = & \Big( M^2 + \frac{\D}{2} \Big)\,\Big( M^2 - \frac{\D}{2} \Big) \, \Bigg[\frac{1}{M^2 - \frac{\D}{2} }-\frac{1}{M^2 + \frac{\D}{2} } \Bigg] \nonumber
 \eea   yields
 \be   \mathbb{G} = \frac{\mathbb{P}_-}{M^2 - \frac{\D}{2} } +  \frac{\mathbb{P}_+}{M^2 + \frac{\D}{2} }~~;~~ \mathbb{P}_\pm = \frac{1}{2} \Big(\mathbf{1}\pm \mathbb{R} \Big) \,,\label{greenfin}   \ee with
 \be \mathbb{R} = \left(
                                                                           \begin{array}{cc}
                                                                             \alpha & \beta \\
                                                                             \gamma & -\alpha \\
                                                                           \end{array}
                                                                         \right)   ~~;~~ \mathbb{R}^2 = \mathbf{1}\,, \label{capR}\ee where the last equality follows from the identity (\ref{idal}), therefore the matrices $\mathbb{P}_\pm $ are projectors, namely
 \be   \mathbb{P}^2_\pm = \mathbb{P}_\pm \,,\label{projectors}\ee
           hence their eigenvalues are $0,1$.

\end{document}